\documentclass[bimj,fleqn]{w-art}
\usepackage{times}
\usepackage{w-thm}
\usepackage[authoryear]{natbib}
\RequirePackage{bm,amsthm,amsmath,amsfonts,amssymb,algorithm,graphicx,changepage,multirow,booktabs,threeparttable,verbatim,subfigure,longtable,mathrsfs}
\usepackage[colorlinks,bookmarksopen,bookmarksnumbered,citecolor=red,urlcolor=red]{hyperref}
\theoremstyle{plain}

\theoremstyle{definition}

\usepackage[]{graphicx}
\chardef\bslash=`\\ 

\begin{document}

\title[Running title]{Regression-based heterogeneity analysis to identify overlapping subgroup structure in high-dimensional data}
\author[First Author]{Ziye Luo\inst{1}}
\author[Second Author]{Xinyue Yao\inst{2}}
\author[Third Author]{Yifan Sun\footnote{Corresponding author: {\sf{e-mail: sunyifan@ruc.edu.cn}}}\inst{1}}
\author[Fourth Author]{Xinyan Fan\footnote{Corresponding author: {\sf{e-mail: 1031820039@qq.com}}}\inst{1}}
\address[\inst{1}]{Center for Applied Statistics, School of Statistics, Renmin University of China, No. 59 Zhongguancun Street, Beijing, 100872, China.}
\address[\inst{2}]{School of Foreign Languages, Renmin University of China, Beijing, China, No. 59 Zhongguancun Street, Beijing, 100872, China.}


\begin{abstract}
Heterogeneity is a hallmark of complex diseases. Regression-based heterogeneity analysis, which is directly concerned with outcome-feature relationships, has led to a deeper understanding of disease biology. Such an analysis identifies the underlying subgroup structure and estimates the subgroup-specific regression coefficients. However, most of the existing regression-based heterogeneity analyses can only address disjoint subgroups; that is, each sample is assigned to only one subgroup. In reality, some samples have multiple labels, for example, many genes have several biological functions, and some cells of pure cell types transition into other types over time, which suggest that their outcome-feature relationships (regression coefficients) can be a mixture of relationships in more than one subgroups, and as a result, the disjoint subgrouping results can be unsatisfactory. To this end, we develop a novel approach to regression-based heterogeneity analysis, which takes into account possible overlaps between subgroups and high data dimensions. A subgroup membership vector is introduced for each sample, which is combined with a loss function. Considering the lack of information arising from small sample sizes, an $l_2$ norm penalty is developed for each membership vector to encourage similarity in its elements. A sparse penalization is also applied for regularized estimation and feature selection. Extensive simulations demonstrate its superiority over direct competitors. The analysis of Cancer Cell Line Encyclopedia data and lung cancer data from The Cancer Genome Atlas shows that the proposed approach can identify an overlapping subgroup structure with favorable performance in prediction and stability. 
\end{abstract}

\maketitle                   




\renewcommand{\leftmark}
 {Ziye Luo et al.: Heterogeneity analysis to identify overlapping subgroup structure}


\section{Introduction}\label{sec1}
Common diseases including cancer are heterogeneous \cite[]{Burrell2013The, Devi2018Heterogeneity}. Various subtypes exist for these diseases, which vary in pathogenesis and prognosis. Heterogeneity of a same disease is challenging and critical in medicine. The high-throughput profiling technologies generate a large amount of high-dimensional molecular data, which have stimulated an increasing number of heterogeneity analyses based on these high-dimensional data. From a methodological perspective, these heterogeneity analyses can be roughly grouped into two classes: clustering-based and regression-based. Clustering-based heterogeneity analysis clusters samples based solely on features, such as diagnostic information, clinical details, and omics data \cite[]{Li2018, Sotiriou2013Breast}. Regression-based heterogeneity analysis is directly concerned with the relationship between outcomes and features. Its target is to identify the subgroup structure and estimate the subgroup-specific regression coefficients \cite[]{He2020}. Such an analysis can lead to a deeper understanding of pathogenic mechanisms, provide a new way to classify diseases, and further facilitate the development of personalized treatments and drugs \cite[]{Dagogo2018Tumour}. 

Many of the existing heterogeneity studies assume disjoint subgroups. In fact, overlapping between subgroups are not rare. Taking clustering as an example. It is well known that many genes have several biological functions and thus may belong to more than one functional subgroups, e.g. pathways \cite[]{Hidalgo2018}. The limitation of disjoint subgroups of genes is especially problematic, ``when many of the genes are likely to be similarly expressed with different groups in response to different subsets of the experiments"\cite[]{Gasch2002}. In cell clustering analysis, as is ubiquitous in nerve cells and cancer cells, not all cells have a specific label, and there are a large number of mixed cells that are transitioning between cell types \cite[]{Kowalczyk2009, Keren2017}. Ignoring the mixture of discrete cell types will make the clustering result tedious and redundant \cite[]{Zhu2019}. Many clustering-based heterogeneity approaches have been developed to overcome the problem of overlapping clusters, and their superiority over disjoint-clustering approaches have been demonstrated in a variety of scenarios. For more details, we refer to Baadel et al.\cite[]{baadel2016overlapping} and Hidalgo et al.\cite[]{Hidalgo2018} However, literature on regression-based heterogeneity analysis is very limited. 
 
\begin{figure}[h]
  \centering
  \includegraphics[width=0.8\textwidth]{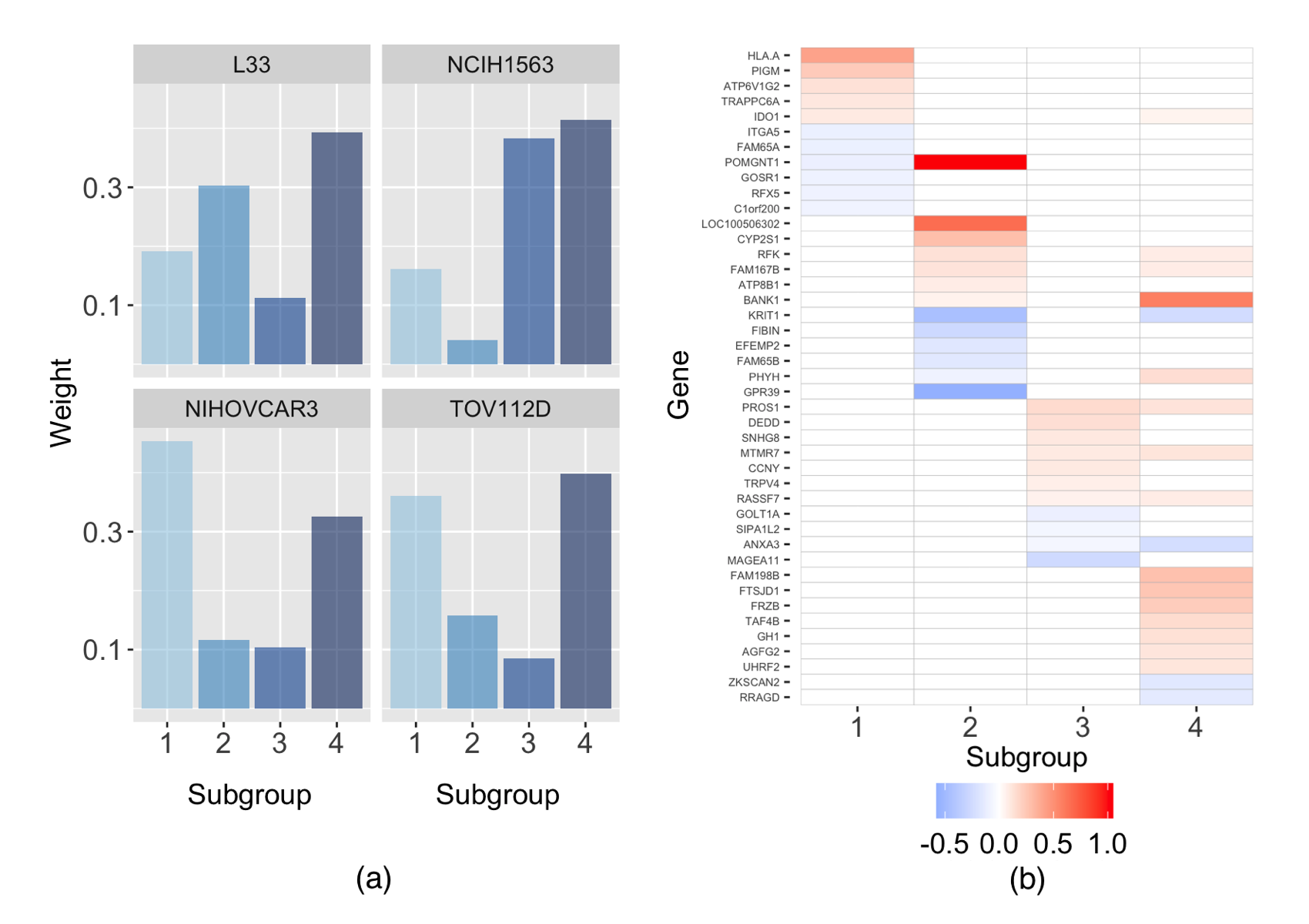}
  \caption{ Analysis of CCLE data. (a): cells that have similar weights in two or more subgroups. The titles are the names of cell lines; 
   (b): heatmap of the estimated regression coefficient matrix for four subgroups (significant variable only). In the heatmaps, the values are represented with different colors, as indicated by the colorbar.}
   \label{fig:toy_example}
\end{figure}

Overlaps between subgroups in terms of outcome-feature relationships are also biomedical relevant. There are many motivating examples, and we provide three here. In the analysis of inpatient length of stay, Wang et al. \cite[]{{Wang2010A}} considered patients belonging to each subgroup with some proportion. In the studies of relapsing polychondritis, Shimizu et al.\cite[]{Jun2018Relapsing} identified some relapsing polychondritis patients who belong to two subgroups, and these patients have distinctive clinical characteristics compared with those patients within only one subgroup.  In the drug sensitivity prediction, consider the mixture of the relationships between gene expressions and responses of the cell to anti-cancer agents. {\color{blue} Specifically}, consider the CCLE (Cancer Cell Line Encyclopedia) data analyzed in study (for more details, refer to Section 4.1). This dataset provides 947 cancer cell lines from nine cancer types with associated gene expression and responses to 24 anti-cancer agents. In Figure \ref{fig:toy_example}, we present the result of PF2341066, one of anti-cancer agents, by using the proposed regression-based heterogeneity analysis approach to be introduced. Four subgroups of cells are identified with distinct patterns of association between gene expressions and response (Figure \ref{fig:toy_example} (b)). It is noted that there are some cells with a maximum membership degree (called weight) below 0.5, indicating that these cells belong to more than one subgroup (Figure \ref{fig:toy_example} (a)). However, these mixed cells can not be detected by using most (if not all) of the aforementioned regression-based heterogeneity analysis approaches. This practical example motivates a need for new regression-based heterogeneity analysis approaches that can accommodate overlapping subgroup structure for practical application in biomedical studies. 

In this paper, we focus on the regression-based heterogeneity analysis, which is challenging because the outcome-feature relationships are not observed directly but can only inferred from the data. There are three classes of strategies to  address this problem. The first one is penalized fusion, which identifies the latent subgroups by penalizing coefficient differences between pairs of samples \cite[]{Ma2017A, Ma2020}. It has multiple advantages, but it also incurs a much larger number of parameters, which causes computational challenges and further reduces the reliability of results, especially when the input data is high-dimensional. The second one is the integration of clustering analysis, such as K-means, into the regression framework, which aims to address the minimization of an objective function over possible groupings by using iterative algorithms\cite[]{Bonhomme2015Grouped, Ando2016, Zhang2019Quantile}. However, the above two strategies cannot handle overlaps between groups. The last class of strategy, and perhaps the most popular strategy, is the finite mixture regression (FMR) model \cite[]{Mclachlan2000Finite} as well as the finite mixture of regression expert (FMRE) model \cite[]{Huynh2019, Chamroukhi2019, Deleforge2015, Devijver2017, Perthame2018}. These models {\color{blue}group} the observations into subgroups by approximating the conditional density of responses given covariates \cite[]{Nguyen2019, Nguyen2020a, Nguyen2020b, Nguyen2020c}. Recently, FMR and FMRE models have been generalized to high-dimensional data via regularization \cite[] {Khalili2007, Khalili2013, Lloyd2018, Stadler2010, Devijver2015, Devijver2017a, Devijver2017b} and other techniques, e.g., inverse regression \cite[]{Nguyen2021a, Nguyen2021b}. Although FMR models are suitable for overlapping  subgroups, they assume that all observations belong to one subgroup with the same probability, which may not reasonable in practice. It is noted that FMRE models relax this assumption and allow each observation has unique mixture probabilities. However, they usually assume that the mixing probabilities depend on the features.

 In this study, we propose a novel regression-based heterogeneity analysis approach. Complementing the existing literature, the proposed approach allows for overlapping subgroups and can be more flexible. Advanced from the penalized fusion and clustering-based methods, sparse penalization is introduced to accommodate high dimensionality and distinguish signals from noises.  Compared to the FMR-related approaches, it has few requirements for the conditional distribution of outcome and more flexible for the proportion of each observation belonging to each subgroup. Our numerical study suggests that the proposed approach has favorable practical performance. With significance advancements in methodology and numerical performance, the proposed approach can provide a powerful new avenue for heterogeneous data analysis.

\section{Methods}
\subsection{Formulation}
Consider a dataset with $n$ independent samples from $K$ subgroups. For the sample $i$, denote $X_i=(x_{i1},x_{i2},\ldots,x_{ip})$ as the $p$-dimensional covariate vector $(p>n)$ and $y_i$ as the outcome. Denote $u_{ki}$ as the probability of sample $i$ belonging to subgroup $k$ with $0\leq u_{ki}\leq 1$ and $\sum_{k=1}^K u_{ki}=1$.  An overlapping subgroup structure is allowed in the sense that a sample can be a member of multiple subgroups with $u_{ki}>0$. For the samples belonging to only one subgroup, their outcomes and covariate vectors obey unique regression equation, that is, 
$$
y_i = X_i\alpha_k+\varepsilon_{i}\ \ \mbox{if}\ u_{ki}=1,
$$
where $\alpha_k=(\alpha_{k1},\alpha_{k2},\ldots,\alpha_{kp})^\top$ is the unique regression coefficients, and $\varepsilon_{i}$'s are random errors. For samples belonging to two or more subgroups, we have $E(y_i)=X_i\sum_{k=1}^{K}u_{ki}\alpha_k$.

We propose a new approach to simultaneously learn the overlapping subgroup structure, identify the important features in each subgroup, and estimate their coefficients. Formally, our approach is formulated as follows: 
\begin{equation}
\label{eq:problem}
\begin{aligned}
 \min_{\bm A, \bm U}\ \ & \sum_{i=1}^{n}\sum_{k=1}^{K}u_{ki}^m(y_i - X_i \alpha_k)^2 + \sum_{k=1}^K\lambda_k \|{\alpha}_k\|_1 + \gamma \sum_{i=1}^n\sum_{k=2}^K(u_{(k),i}-u_{(k-1),i})^2 \\
\text{s.t.}\ \ & \sum_{k=1}^K u_{ki} = 1, \ 0 \leq u_{ki} \leq 1.\\
 \end{aligned}
 \end{equation}
 where $u_{(1),i}\leq u_{(2),i}\leq \dots u_{(K),i}$ are the ordered elements of $(u_{1i},\ldots,u_{Ki})^{\top}$, $m$ is a constant, and  $\lambda\geq 0, \gamma\geq 0$ are two tuning parameters. Denote 
 $\bm A=(\alpha_1,\alpha_2,\ldots,\alpha_K)$ as the $p\times K$ coefficient matrix that collects together all the regression coefficients of $K$ subgroups, and $\bm U=(u_1,u_2,\ldots,u_n)$ as the $K\times n$ weight matrix with $u_i=(u_{1i},u_{2i},\ldots,u_{Ki})^\top$.



In the objective function in Equation (\ref{eq:problem}), the first term measures the lack-of-fit in the form of weighted squared loss. The weighted strategy is motivated by the kernel smoothing method for local linear regression. The functions of $u_{ik}$'s take the place of kernels. The larger the probability of sample $i$ belonging to subgroup $k$ is, the more important the lack-of-fit between $y_i$ and $X_i \alpha_k$ is. In the special case $u_{ki}\in\{0,1\}$, the first term recovers the measure in the heterogeneity analysis that integrates K-means into the regression framework \cite[]{Bonhomme2015Grouped, Zhang2019Quantile}. Also, the weighted squared loss can be found in fuzzy clustering. In the loss function, $m$ can be any real number, and following the suggestion of Pal and Bezdek\cite{Pal1995On}, we set $m=2$. Compared to the existing regression-based heterogeneity analysis, which partition samples into disjoint subgroups, the proposed approach allows a sample to belong to multiple subgroups, and thus can lead to more flexible grouping results. 

The first penalty of Equation (\ref{eq:problem}) is Lasso, which is responsible for feature selection and sparse estimation. The parameters $\lambda_k$'s allow for the possibility of controlling the extent of sparsity for a specific subgroup. A larger $\lambda_k$ makes the corresponding $\alpha_k$ sparser. Lasso can be replaced by more complicated penalties, for example, MCP and SCAD.  The main advancement of proposed method is the second penalty, which penalizes the distance between successive weights (sorted in ascending order) in a similar way as is done in overlapping clustering analysis \cite[]{Hidalgo2018}. It is minimized when $u_{ki}=1/K$; thus, the weights tend to move away from the extreme values (i.e., $u_{ki}\in\{0,1\}$) and toward one another. In this regard, this penalty is specially designed to accommodate the overlapping subgroup structure. Different from the negative-entropy-type penalty used in regularized fuzzy clustering \cite[]{Suk2010Regularized}, it does not force each sample to belong to multiple subgroups, and is therefore more flexible. This penalty takes a conservative strategy in the sense that it does not assign a sample to a specific subgroup unless there is sufficient evidence. Considering that insufficient information can arise from a small sample size or other reasons, this conservative strategy is a reasonable choice. Another advantage of this penalty is that the quadratic form makes it computationally easier than the absolute value-based form.

Denote $(\hat{\bm U},\hat{\bm A})$ as the solution of optimization problem (\ref{eq:problem}). The subgroup membership of $n$ samples are reflected in weight matrix $\hat{\bm U}$. When $\hat u_{1i},\hat u_{2i},\ldots, \hat u_{Ki}$ are close to each other, there is insufficient evidence to determine the unique subgroup membership for the $i$th sample; {\color{blue} on the other hand, when there is one weight that is much larger than others, there is dominating evidence to assign the $i$th sample to a specific subgroup}. In practice, if the subgroup membership needs to be specified, we can simply assign each sample to the subgroup whose weight is the largest. The estimated coefficients of $K$ subgroups are reflected in matrix $\hat{\bm A}$. A nonzero component corresponds to a feature that is associated with the response. 


\subsection{Computation}
The optimization problem in Equation (\ref{eq:problem}) is bi-convex with respect to coefficient matrix $\bm A$ and weight matrix $\bm U$; that is, for a given $\bm A$, it is convex for $\bm U$, and vice versa. As such, we apply an alternating optimization to obtain the partial minimum of the objective function. Specifically, it starts from an initial estimate of $\bm U$, and then updates $\bm A$ and $\bm U$ sequentially until the convergence is reached. 

\subsubsection{Initial estimate}
\label{sec:initial}
Since the objective function in Equation (\ref{eq:problem}) is non-convex, the initial estimate of matrix $\bm U$ is crucial to the quality of the solution. In the existing $K$-means-based subgroup analysis, a common practice is to draw multiple initial estimates randomly and select the one yielding the lowest objective function value. This practice has been shown to work well for data with modest size. However, the required number of initial estimates soars along with the increase of $p$, making this practice unfeasible for large-scale data. An alternative is the variable neighborhood search method (see more details in Appendix), but computing high-dimensional data using this method is currently challenging (see more discussion in Section \ref{sec:sim}).  

As such, we propose a novel method to obtain a reasonable initial estimate of $\bm U$. This method is based on a delicate feature screening procedure. As shown in Figure \ref{fig:workflow}, the procedure for setting initial values involves the following steps: 

\begin{figure}[H]
  \centering
  \includegraphics[width=0.5\textwidth]{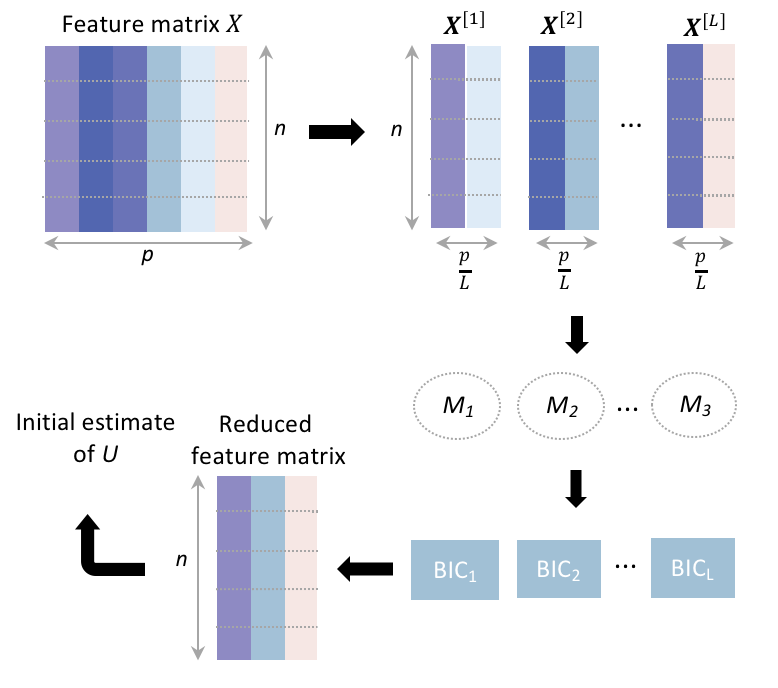}
  \caption{A diagram illustrating the workflow for generating the initial estimate of $\bm U$.}
   \label{fig:workflow}
\end{figure}

\noindent {\bf Step 1: Partition.} Partition $p$ features into $L$ non-overlapping subsets with equal size. Denote $[l]$ as the index set of $l$th subset of features. For a $p$-dimensional vector $z=(z_1,\ldots,z_p)^\top$, let $z^{[l]}$ denote its sub-vector indexed by set $[l]$. The full dataset is thus partitioned into $L$ subsets $\{X_i^{[1]},y_i\}_{i=1}^n, \ldots, \{X_i^{[L]},y_i\}_{i=1}^n$.  

\noindent {\bf Step 2: Estimate.} {\color{blue} For the $l$th data subset, consider the submodel $M_l$:
\[
M_l: y_i = X_{i}^{[l]}{\theta}_{k}^{[l]}+ \varepsilon_{i},
\]
where ${\theta}_{k}^{[l]}$ is a $\lfloor\frac{p}{L}\rfloor$-dimensional coefficient vector and $\varepsilon_{i}$ is the error.  In the submodel $M_l$, the $n$ samples are assumed to be partitioned into $K_0$ disjoint subgroups,} and each subgroup $k$ has unique regression coefficients $\theta_{k}^{[l]}$. Because different data subsets have different features, different data subsets may have distinct subgroup structures as well as different subgroup-specific regression coefficients. Note that $K_0$ is not required to be the true number of subgroups $K$. {\color{blue} Since the results, i.e., the initial estimates of $\bm U$, are not sensitive to the choice of $K_0$, we recommend using a relatively small $K_0$ in practice to reduce the computation cost. Throughout our numerical studies, $K_0$ is set to be 2.} To determine the grouping and subgroup-specific coefficients, we solve the following optimization problem: 
\begin{equation}
\label{eq:initial}
\begin{aligned}
&\min_{\{w_{ki}^{[l]}\},\{\theta_{k}^{[l]}\}} \sum_{i=1}^{n}\sum_{k=1}^{K_0}w_{ki}^{[l]}(y_i - X_{i}^{[l]} \theta_{k}^{[l]})^2 + \sum_{k=1}^{K_0}\lambda_k \|{\theta}_{k}^{[l]}\|_1 \\
&\text{s.t.}\ \ \sum_{k=1}^{K_0} w_{ki}^{[l]} = 1, \ w_{ki}^{[l]}\in\{0,1\}.\\
 \end{aligned}
 \end{equation} 
where the weight $w_{ki}^{[l]}$ is a binary variable: $w_{ki}^{[l]}=1$ if $i$th sample belongs to $k$th subgroup in $l$th data subset , and $w_{ki}^{[l]}=0$ otherwise. This optimization problem can be solved via a simple iterative strategy, which is summarized in Algorithm \ref{alg1}. 
\begin{algorithm}
\caption{Iterative strategy to solve optimization problem (\ref{eq:initial})}\label{alg1}
1. Initialization. Set $t=0$. Randomly assign $n$ samples into $K_0$ disjoint subgroups and initialize the weights $w_{ki}^{(t)}$'s. 

2. Update. For $k\in \{1,2,\ldots,K_0\}$, compute $\theta_k^{(t+1)}=\text{argmin}_{\theta}\sum_{i=1}^n w_{ki}^{(t)} (y_i-X_i\theta)^2+\lambda_k \|{\theta}\|_1$.

3. Assignment. For $i\in\{1,2,\ldots,n\}$, compute $k^\star=\text{argmin}_{k\in\{1,2,\ldots,K_0\}}(y_i-X_i \theta_k^{(t+1)})^2$. Set $w_{k^\star i}^{(t+1)}=1$, and $w_{ki}^{(t+1)}=0$ for $k\neq k^\star$. 

4. Set $t=t+1$ and go to Step 2 until the objective function in Equation (\ref{eq:initial}) converges. 
\end{algorithm}
Denote the estimate as $\{\{\hat{w}_{ki}^{[l]}\}_{i=1}^n,\hat{\theta}_k^{[l]}\}_{k=1}^{K_0}$. The $L$ sets of estimates, as opposed to the individual subgrouping results, will be used in the following step. 

\noindent {\bf Step 3: Select.} For each submodel $M_l$, we calculate its Bayes Information Criterion (BIC), which is defined as follows: 
{\color{blue}
\begin{equation}
\label{eq:BIC1}
\text{BIC}_l=\log\Big[\sum_{i=1}^n\sum_{k=1}^{K_0}\hat{w}_{ki}^{[l]}(y_i-X_{i}^{[l]}\hat{\theta}_{k}^{[l]})^2/n\Big]+df_l\log n/n, 
\end{equation}
}  
where $df_l=\sum_{j=1}^{p/L}\sum_{k=1}^{K_0}\mathbb{I}_{\{\hat{\theta}_{kj}^{[l]}\neq 0\}}$ is the number of nonzero regression coefficients, and $\mathbb{I}_{\{\hat{\theta}_{kj}^{[l]}\neq 0\}}$ is the indicator function that takes value 1 when $\hat{\theta}_{kj}^{[l]}\neq 0$, and value 0 otherwise. After sorting $L$ BIC values in an ascending order, the first $s$ submodels in the ranked list are kept, and all the features with nonzero coefficients in the $s$ submodels are selected. Denote $A_s$ as the index of selected features.

\noindent {\bf Step 4: Re-estimate.} For data subset $\{X_i^{A_s},y_i\}_{i=1}^n$, repeat the Estimation step (Step 2) where the samples are partitioned into $K$ 
disjoint subgroups. The initial estimate of weight matrix $\bm{U}$ is given by the output $(\hat{w}_{ki})_{K\times n}$. 
 
\noindent{\bf Remarks} 
In practice, we need to choose the total number of submodels $L$ and the number of selected submodels $s$. Based on our experience, our initialization method performs similarly when $L$ is relatively large such that each submodel has a small number (5 or 10, for example) of features. To obtain a robust initial estimate of the weight matrix, we select $s$ such that there are 30--50 nonzero coefficients in these submodels.

\subsubsection{Updated coefficient matrix}

For fixed weight matrix $\bm U$, the objective function in Equation (\ref{eq:problem}) with respect to coefficient matrix $\bm A$, becomes as follows: 
\begin{equation*}
\sum_{i=1}^n \sum_{k=1}^K u_{ki}^m(y_i-X_i\alpha_k)^2+\sum_{k=1}^K \lambda_k \|\alpha_k\|_1.
\end{equation*} 
This function is separable with respect to $\alpha_k$ ($k=1,2,\ldots, K$), and thus each $\alpha_k$ can be solved separately: 
\[\hat{\alpha}_k=\text{argmin}_{\alpha} \sum_{i=1}^nu_{ki}^m(y_i-X_i\alpha)^2+\lambda_k\|\alpha\|_1.\]
Introduce ${\tilde y}_{ki}=u_{ki}^{m/2}y_i$ and ${\tilde X}_{ki}=u_{ki}^{m/2}X_i$. Then, the above optimization problem 
can be reformulated as a standard lasso problem: 
\[\hat{\alpha}_k=\text{argmin}_{\alpha} \sum_{i=1}^n({\tilde y}_{ik}-{\tilde X}_{ik}\alpha)^2+\lambda_k\|\alpha\|_1,\]
and as a result, many available efficient algorithms can be utilized. In the numerical study, we adopt the coordinate decent algorithm, which is a well-developed algorithm for tackling high-dimensional data. 

\subsubsection{Updated weight matrix}
Given coefficient matrix $\bm A$, the optimization problem with respect to weight matrix $\bm U$ is as follows: 
\begin{equation*}
\begin{aligned}
 \min_{\bm U}\ \ & \sum_{i=1}^{n}\sum_{k=1}^{K}u_{ki}^m(y_i - X_i \alpha_k)^2 + \gamma \sum_{i=1}^n\sum_{k=2}^K (u_{(k),i}-u_{(k-1),i})^2 \\
\text{s.t.}\ \ & \sum_{k=1}^K u_{ki} = 1,\ 0 \leq u_{ki} \leq 1.\\
 \end{aligned}
 \end{equation*} 
This procedure is learning the underlying grouping structure and can be solved by the interior-point method. The detailed calculation is provided in the Appendix.  

Denote $\bm U^{(t)}$ and $\bm A^{(t)}$ as the estimate of $\bm U$ and $\bm A$ in the $t$th iteration, respectively. The overall procedure of the proposed approach are summarized in Algorithm \ref{alg2}. 

\begin{algorithm}
\caption{Alternating optimization of the objective function in Equation (1)}\label{alg2}

1. Initialization. Set $t=0$. Estimate an initial weight matrix $\bm U^{(0)}$. 

2. Update coefficient matrix $\bm A^{(t+1)}$.
 
3. Update weight matrix $\bm U^{(t+1)}$. 

4. Set $t=t+1$. Repeat Steps 2--3 until the objective function of Equation (1) converges. 
\end{algorithm}

\noindent{\bf Tuning parameter selection} 
The proposed approach involves tuning the parameters: regularization parameters $\lambda_k$'s and $\gamma$ as well as the number of subgroups $K$. The values of $\lambda_k$'s are determined by five-fold cross validation (CV) during the update of $\bm A$ in Algorithm 1. Standard 
tuning parameter selection is conducted following the literature \cite[]{Friedman2010}. For $\gamma$, we chose its optimal value using five-fold CV with $\gamma \in [0.01, 0.1, 0.5, 1, 5, 10]$. In our numerical study, we use the modified BIC \cite[]{Stadler2010} for heterogeneous high-dimensional data settings to select $K$ by minimizing 

\begin{equation}
\label{eq:bic}
\text{BIC}=\log\Big[\sum_{i=1}^n\sum_{k=1}^K\hat u_{ki}^m(y_i-X_i\hat  \alpha_k)^2/n\Big]+d_f\log n/n,
\end{equation}
where $d_f=K+(K-1)+\sum_{j=1}^p\sum_{k=1}^K\mathbb{I}_{\{\hat {\alpha}_{kj}\neq 0\}}$. We acknowledge the importance of selection of $K$ (including optimality, sensitivity, etc.). As BIC has been extensively adopted in the literature, we choose not to discuss further.

\noindent{\bf Realization} 
To facilitate data analysis within and beyond this study, we have developed a Python code implementing the proposed approach and made it publicly available at https://github.com/foliag/subgroup. The proposed approach is computationally affordable. For example, with a fixed number of subgroups $K$, for a simulated dataset with 200 samples divided into two disjoint subgroups, 1000 features, and three important features in each subgroup, the analysis can be accomplished within 200 seconds using a laptop with standard configurations. In addition, the computational time of the proposed approach is linear, with an increasing number of features (Figure \ref{fig:time}), and thus, it is suitable for large-scale data. 

\section{Simulation}
\label{sec:sim}
This section aims to assess the performance of the proposed approach under a wide spectrum of simulation settings. The following five alternatives are used for comparison: (a) L-MLR, a lasso-penalized mixture of linear regression models \cite[]{Lloyd2018}; (b) MoE, a mixtures-of-experts model for high-dimensional data \cite[]{Huynh2019}; (c) S-FMR, which advances from existing finite-mixture-regression methods by focusing on the structure of covariate effects \cite[]{liu2020}; (d) FCM1, which first uses the standard fuzzy C-means method to cluster the samples based on response variable and $p$ features, and then applies this to each subgroup; and (e) FCM2, which applies Lasso again to the whole dataset under the homogeneity assumption (i.e., $y_i=X_i\alpha+\varepsilon_i$), clusters samples based on the residuals $y_i-X_i\alpha$ via the fuzzy C-means method, and then applies Lasso to each subgroup. 
{\color{blue} For all alternatives, we give the true number of subgroups as input. }

For the evaluation of grouping performance, we report the estimated number of subgroups ($\hat{K}$) and percentage of $\hat{K}$ equaling the true number of subgroups. We compute the true positive rates (TPR) and false positive rates (FPR), which are averaged over subgroups, to measure the feature selection accuracy. Estimation accuracy is measured by root mean squared errors (RMSE), defined as $\sqrt{\frac{1}{pK} \sum^K_{k=1}\sum^p_{j=1}\|\hat{\alpha}_{ki}-\alpha_{ki}\|^2_2}$, and prediction accuracy is measured by root prediction errors (RPE), defined as $\sqrt{\frac{1}{n}\sum^n_{i=1}\big(\hat E(y_i)-E(y_i)\big)^2}$, where $\hat E(y_i)=X_i\sum^K_{k=1}\hat u_{ki}\hat\alpha_k$. {\color{blue}To quantify the accuracy of the grouping, we use the average $L_1$ loss per sample and the adjusted Rand index (ARI). The average $L_1$ loss per sample (i.e., $\|\hat{\bm U}-\bm U\|_1/n$) measures the average difference between the estimated weight matrix and true weight matrix, where $\|\cdot \|$ is the usual vector $L_1$ norm after vectorization. The lower $L_1$ loss indicates better subgrouping performance. ARI is a standard index in clustering analysis used to assess the consistency between the estimated group structure and the true group structure. }

We simulate heterogeneous data with $p=1000$ features and $n$ samples with $n=200$ and 400. The $p$ features are generated from a multivariate normal distribution with marginal mean 0 and variance 1. We consider an auto-regressive (AR) correlation structure, where $j$th and $k$th features have the correlation coefficient $0.5^{|j-k|}$. The response variables are generated from the heterogeneous regression model, of which the random errors are independently generated from normal distributions with mean 0 and standard deviation $\sigma=0.5$ and $1$. 

\noindent{\bf Example 1} \rm (Overlapping subgroups). The $n$ samples are divided into two overlapping subgroups. There are $n/5$ samples belonging to two subgroups with distinct membership weights $a_i$'s. Each $a_i$ is generated from uniform distribution $\mathcal{U}[0,1]$:
 
\[\bm U=\left(
\begin{array}{cc}
1_{\frac{2n}{5}\times 1} & 0_{\frac{2n}{5}\times 1} \\
0_{\frac{2n}{5}\times 1} & 1_{\frac{2n}{5}\times 1} \\
a_1 & 1-a_1 \\
\vdots & \vdots \\
a_{\frac{n}{5}} & 1-a_{\frac{n}{5}} \\
\end{array}
\right).\]
For the true coefficient vectors, we consider two scenarios.

\noindent (S1)
\[\bm A=\left(
\begin{array}{cc}
1 & 0 \\
2 & 0 \\
3 & 0 \\
0 & -4 \\
0 & -5 \\
0 & -6 \\
0_{(p-6)\times 1} & 0_{(p-6)\times 1} \\
\end{array}
\right).\]
The two coefficient vectors have distinct sparsity structures.

\noindent (S2)
\[\bm A=\left(
\begin{array}{cc}
1 & 1 \\
2 & -2 \\
3 & -3 \\
0_{(p-3)\times 1} & 0_{(p-3)\times 1} \\
\end{array}
\right).\]
The two coefficient vectors have the same sparsity structure. The first coefficient is homogeneous across two subgroups.

With the proposed approach, we first examine the values of BIC as a function of number of subgroups $K$. Figure \ref{bic_simu} presents the BIC curve for a random replicate under setting S1, $n=200$, and $\sigma=0.5$. The optimal point with $K=2$ is clearly identified. We also examined a few other replicates and observed similar patterns. We then compute summary statistics based on 100 replicates. Table \ref{kselect_example1e5pr5} reports the sample mean, median, and standard deviation of the estimated number of subgroups $\hat{K}$ as well as the percentage that $\hat{K}$ equals to the true number of subgroups by the proposed approach with $\sigma=0.5$. The median of $\hat K$ is 2, which is the true number of subgroups for all settings. As $n$ increases, the mean reaches closer to 2, the standard deviation becomes smaller, and the percentage of correctly determining the number of subgroups becomes larger. 

\begin{table}
\centering
\caption{Simulation results for Example 1: the sample mean, median and standard deviation (s.d.) of $\hat{K}$ and the percentage 
of $\hat{K}$ equaling to the true number of subgroups by the proposed approach based on 100 replicates with $\sigma=0.5$.}
\label{kselect_example1e5pr5}
\begin{tabular}{@{}ccccccccc@{}}
\toprule
             & \multicolumn{4}{c}{$n=200$} & \multicolumn{4}{c}{$n=400$}  \\ 
             \cmidrule(r){2-5}\cmidrule(r){6-9}
             & Mean & Median & s.d. & Percentage & Mean & Median & s.d. & Percentage \\
             \midrule
S1 & 2.03                     & 2.00           & 0.17         & 0.96           & 2.03                     & 2.00                & 0.16                           & 0.97                    \\ 
S2 & 2.02                     & 2.00           & 0.16         & 0.97           & 2.02                  & 2.00                & 0.15                           & 0.98                    \\ \bottomrule
\end{tabular}
\end{table}

\begin{figure}[htb]
   \centering
   \includegraphics[width=0.95\textwidth]{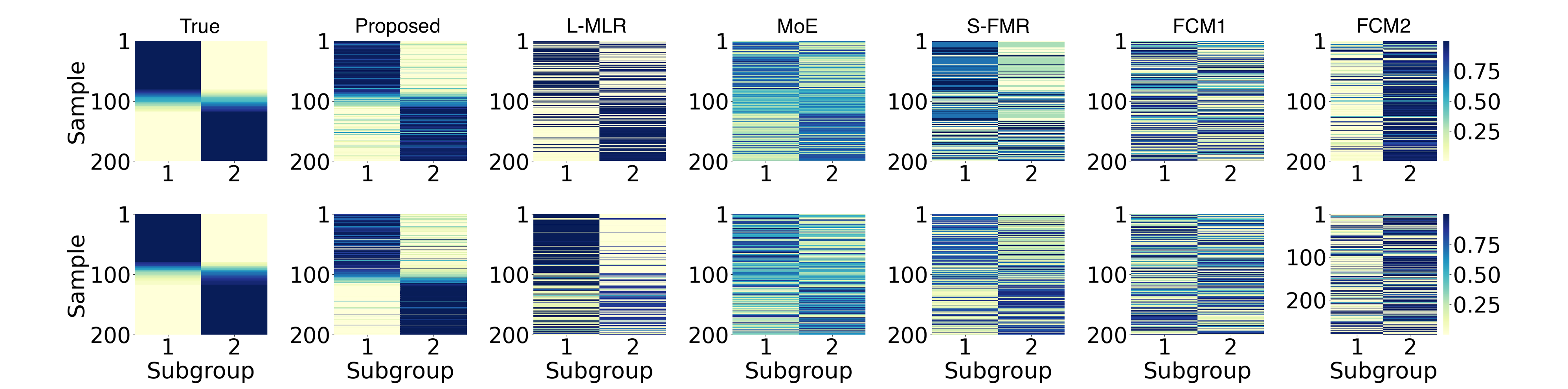}
   \caption{Simulation results for Example 1: heatmaps of true and estimated weight matrix for one replicate with $n=200$ and $\sigma=0.5$. The weights are represented with different colors, as indicated by the colorbar. Top: setting S1. Bottom: setting S2. }
   \label{fig:ex1n2_weight_heatmap}
\end{figure}

In the following, we give the true $K$ as input to all methods. Figure \ref{fig:ex1n2_weight_heatmap} presents the true and estimated weight matrices for one replicate with $n=200$ and $\sigma=0.5$. We can observe that only the proposed approach can identify the underlying subgroup structure. {\color{blue} In addition, the proposed approach achieves the best performance in terms of the average $L_1$ loss under all scenarios (Figure \ref{fig:l1loss_ex1e5} and \ref{fig:l1loss_ex1e1})}, and the $L_1$ loss of the proposed approach becomes lower as $n$ increases.

Table \ref{tab:ex1e5} presents the TPR, FPR, RMSE, and RPE of the six methods with $\sigma=0.5$. The values in the parentheses are the standard deviations of the corresponding metrics values. The proposed approach has superior performance in feature selection, as it can identify more important features while having a low number of false positives. In terms of estimation and prediction, the proposed approach also outperforms the alternatives. In all the scenarios considered here, the proposed approach has the lowest estimation errors and prediction errors. Similar observations are also made under a larger random error, as presented in the Table \ref{tab:ex1e1}. 

We conduct {\color{blue} an} additional simulation to examine the dependence of the proposed approach on the values of coefficients. Specifically, the coefficient matrix $\bm A$ is 
\[\bm A=\left(
\begin{array}{cc}
1 & 0 \\
1.5 & 0 \\
2 & 0 \\
0 & -1 \\
0 & -1.5 \\
0 & -2 \\
0_{(p-6)\times 1} & 0_{(p-6)\times 1} \\
\end{array}
\right).\]
The other settings are the same as described above. The results with $n=400$ and $\sigma=0.5$ are presented in Table \ref{tab:low_beta1}. The proposed approach is again observed to have favorable performance in subgrouping, feature selection, estimation, and prediction. 

We also consider {\color{blue} the case of $K=4$}. Specifically, $n=900$ samples form four overlapping subgroups and 
\[\bm U=\left(
\begin{array}{cccc}
1_{\frac{n}{5}\times 1} & 0_{\frac{n}{5}\times 1} & 0_{\frac{n}{5}\times 1} & 0_{\frac{n}{5}\times 1}\\
0_{\frac{n}{5}\times 1} & 1_{\frac{n}{5}\times 1} & 0_{\frac{n}{5}\times 1} & 0_{\frac{n}{5}\times 1}\\
0_{\frac{n}{5}\times 1} & 0_{\frac{n}{5}\times 1} & 1_{\frac{n}{5}\times 1} & 0_{\frac{n}{5}\times 1}\\
0_{\frac{n}{5}\times 1} & 0_{\frac{n}{5}\times 1} & 0_{\frac{n}{5}\times 1} & 1_{\frac{n}{5}\times 1}\\
a_1 & 1-a_1 & 0 & 0\\
\vdots & \vdots & \vdots & \vdots \\
a_{\frac{n}{15}} & 1-a_{\frac{n}{15}} & 0 & 0 \\
0 & 0 & a_{\frac{n}{15}+1} & 1-a_{\frac{n}{15}+1} \\
\vdots & \vdots & \vdots & \vdots \\
0 & 0 & a_{\frac{2n}{15}} & 1-a_{\frac{2n}{15}} \\
\tilde{a}_{\frac{2n}{15}+1,1} & \tilde{a}_{\frac{2n}{15}+1,2} & \tilde{a}_{\frac{2n}{15}+1,3} & \tilde{a}_{\frac{2n}{15}+1,4} \\
\vdots & \vdots & \vdots & \vdots \\
\tilde{a}_{\frac{n}{5},1} & \tilde{a}_{\frac{n}{5},2} & \tilde{a}_{\frac{n}{5},3} & \tilde{a}_{\frac{n}{5},4} \\
\end{array}
\right),\]
where each $a_i$ as well as $\tilde{a}_{i,k}$ is generated from uniform distribution $\mathcal{U}[0,1]$, and $\sum_{k=1}^4 \tilde{a}_{i,k}=1$. 
There are $n/15$ samples belonging to both subgroups 1 and 2, $n/15$ samples belonging to both subgroups 3 and 4, and another $n/15$ samples belonging to four subgroups. The true coefficient vectors are as follows:
\begin{equation}
\label{eq:coef}
\bm A=\left(
\begin{array}{cccc}
2 & 0 & 2 & 2 \\
2 & 0 & 0 & 2 \\
3 & 0 & 3 & 0 \\
3 & 0 & 0 & 0 \\
0 & -2 & 0 & 0 \\
0 & -2 & -2 & 0 \\
0 & -3 & 0 & -3\\
0 & -3 & -3 & -3\\
0_{(p-8)\times 1} & 0_{(p-8)\times 1} & 0_{(p-8)\times 1} & 0_{(p-8)\times 1}\\
\end{array}
\right).
\end{equation}
Table \ref{tab:1k4kselect} reports the statistics of estimated number of subgroups $\hat{K}$. The results suggest that the proposed approach still can 
accurately determine the number of subgroups. Other summary statistics are presented in Table \ref{tab:4k}. 
Overall, the performances of all approaches are worse than that under $K=2$. This is as expected since the subgrouping structure under $K=4$ is more complex. However, the proposed approach still outperforms the alternatives.

\begin{table}
{\tiny
\centering
\caption{Simulation results for Example 1: TPR, FPR, RMSE, RPE based on 100 replicates with $\sigma=0.5$. Each cell shows the mean (s.d.)}
\label{tab:ex1e5}
\resizebox{\textwidth}{!}{%
\begin{tabular}{cclllll}
\hline
                     & $n$                  & \multicolumn{1}{c}{Method} & \multicolumn{1}{c}{TPR} & \multicolumn{1}{c}{FPR} & \multicolumn{1}{c}{RMSE} & \multicolumn{1}{c}{RPE} \\ \hline
\multirow{12}{*}{S1} & \multirow{6}{*}{200} & Proposed                    & \textbf{0.903(0.087)}   & \textbf{0.001(0.002)}   & \textbf{0.036(0.010)}    & \textbf{1.247(0.391)}   \\
                     &                      & L-MLR                      & 0.405(0.139)            & 0.138(0.013)            & 0.076(0.008)             & 2.983(0.609)            \\
                     &                      & MoE                        & 0.357(0.045)            & 0.002(0.001)            & 0.090(0.010)             & 4.184(0.557)            \\
                     &                      & S-FMR                      & 0.145(0.176)            & 0.125(0.017)            & 0.129(0.011)             & 3.059(0.822)            \\
                     &                      & FCM1                       & 0.033(0.056)            & 0.006(0.008)            & 0.194(0.008)             & 7.981(0.578)            \\
                     &                      & FCM2                       & 0.053(0.094)            & 0.007(0.009)            & 0.187(0.012)             & 7.003(0.994)            \\ \cline{2-7} 
                     & \multirow{6}{*}{400} & Proposed                    & \textbf{0.910(0.075)}   & \textbf{0.001(0.003)}   & \textbf{0.025(0.003)}    & \textbf{1.056(0.270)}   \\
                     &                      & L-MLR                      & 0.432(0.133)            & 0.124(0.008)            & 0.066(0.013)             & 2.733(0.518)            \\
                     &                      & MoE                        & 0.410(0.068)            & 0.008(0.005)            & 0.066(0.022)             & 3.645(2.649)            \\
                     &                      & S-FMR                      & 0.183(0.160)            & 0.074(0.021)            & 0.117(0.021)             & 2.033(1.053)            \\
                     &                      & FCM1                       & 0.037(0.084)            & 0.007(0.009)            & 0.188(0.008)             & 7.947(0.453)            \\
                     &                      & FCM2                       & 0.093(0.109)            & 0.010(0.013)            & 0.178(0.014)             & 7.345(0.925)            \\ \hline
\multirow{12}{*}{S2} & \multirow{6}{*}{200} & Proposed                    & \textbf{0.875(0.072)}   & \textbf{0.001(0.002)}   & \textbf{0.043(0.008)}    & \textbf{1.396(0.232)}   \\
                     &                      & L-MLR                      & 0.408(0.102)            & 0.132(0.009)            & 0.084(0.009)             & 2.858(0.637)            \\
                     &                      & MoE                        & 0.374(0.047)            & 0.002(0.001)            & 0.116(0.017)             & 4.152(0.654)            \\
                     &                      & S-FMR                      & 0.112(0.006)            & 0.273(0.114)            & 0.130(0.004)             & 2.849(0.134)            \\
                     &                      & FCM1                       & 0.021(0.054)            & \textbf{0.001(0.001)}   & 0.119(0.002)             & 4.148(0.256)            \\
                     &                      & FCM2                       & 0.053(0.093)            & 0.002(0.005)            & 0.122(0.009)             & 3.983(0.470)            \\ \cline{2-7} 
                     & \multirow{6}{*}{400}                   & Proposed                    & \textbf{0.883(0.174)}   & 0.005(0.018)            & \textbf{0.034(0.024)}    & \textbf{1.287(0.129)}   \\
                     &                      & L-MLR                      & 0.413(0.122)            & 0.125(0.007)            & 0.065(0.011)             & 2.708(0.540)            \\
                     &                      & MoE                        & 0.396(0.157)            & 0.006(0.004)            & 0.071(0.056)             & 2.868(1.933)            \\
                     &                      & S-FMR                      & 0.128(0.017)            & 0.225(0.131)            & 0.119(0.009)             & 2.170(0.064)            \\
                     &                      & FCM1                       & 0.030(0.086)            & \textbf{0.001(0.004)}   & 0.082(0.002)             & 4.184(0.161)            \\
                     &                      & FCM2                       & 0.087(0.090)            & \textbf{0.001(0.003)}   & 0.085(0.003)             & 4.116(0.273)            \\ \hline
\end{tabular}%
}}
\end{table}

Although the proposed approach aims to recover the full weight matrix $\bm U$, it can also be used as a non-overlapping subgroup identification method; that is, it can cluster $n$ samples into a series of disjoint subgroups, by labeling each sample as the majority type. We test the proposed approach as a non-overlapping subgroup identification method in the following example. 

\noindent{\bf Example 2} \rm (Non-overlapping subgroups). The $n$ samples are divided into two disjoint subgroups. We consider two types of subgroup structure: (a) balanced, where the two subgroups have the same size, and (b) unbalanced, where the two subgroups have sizes in ratios of 3:7. For the true coefficient vectors, we consider two scenarios. 

\noindent (S3) 
\[\bm A=\left(
\begin{array}{cc}
1 & 1 \\
2 & -2 \\
3 & -3 \\
0_{(p-3)\times 1} & 0_{(p-3)\times 1} \\
\end{array}
\right)\]
The two coefficient vectors have the same sparsity structure. The first coefficient is homogeneous across two subgroups.

\noindent (S4) 
\[\bm A=\left(
\begin{array}{cc}
1 & 0 \\
2 & 0 \\
3 & 0 \\
0 & -1 \\
0 & -2 \\
0 & -3 \\
0_{(p-6)\times 1} & 0_{(p-6)\times 1} \\
\end{array}
\right)\]
The two coefficient vectors have distinct sparsity structures. 

Table \ref{kselect_example2e5pr5} reports the mean, median, and standard deviation of $\hat{K}$ as well as the percentage that $\hat{K}=K$ by the proposed approach, based on 100 replicates for balanced cases with $\sigma=0.5$. We obtain similar patterns as those in Table \ref{kselect_example1e5pr5} for Example 1. Again, the proposed approach can accurately identify the true number of subgroups. 
Figure \ref{fig:ari_ex2e5pr5} displays the results of ARI. The proposed approach is always the highest; therefore, it outperforms alternatives in accuracy of grouping. Table \ref{tab:ex2e5} summarizes the results of other statistics. Similar to Example 1, the proposed approach gives the largest TPR and the lowest FPR among all six methods. In the evaluation of estimation and prediction, the proposed approach is again observed to have favorable performance. The results of other settings are presented in Table \ref{tab:ex2e1}-\ref{tab:ex2e1pr7} and Figure \ref{fig:ari_ex2e1pr5}, and the observed patterns are similar.  We also examine the performance of the proposed approach for smaller coefficients case. Specifically, we use $\bm{A}$ defined in Equation (\ref{eq:coef}) as the coefficient matrix and present the results in Table \ref{tab:low_beta2}. Similar patterns have been observed.

\begin{table}
\centering
\caption{Simulation results: the sample mean, median, and standard deviation (s.d.) of $\hat{K}$ and the percentage 
of $\hat{K}$ equaling the true number of subgroups by the proposed approach based on 100 replicates for balanced cases with $\sigma=0.5$.}
\label{kselect_example2e5pr5}
\begin{tabular}{@{}ccccccccc@{}}
\toprule
             & \multicolumn{4}{c}{$n=200$} & \multicolumn{4}{c}{$n=400$}  \\ 
             \cmidrule(r){2-5}\cmidrule(r){6-9}
             & Mean & Median & s.d. & Percentage & Mean & Median & s.d. & Percentage \\
             \midrule
S3 & 2.05                     & 2.00           &0.22          & 0.95         & 2.15               & 2.00           & 0.37          & 0.85                    \\ 
S4 & 2.03                     & 2.00           & 0.17        & 0.97          & 2.02               & 2.00           & 0.13          & 0.98                    \\ 
\bottomrule
\end{tabular}
\end{table}

\begin{table}[h]
{\tiny
\centering
\caption{Simulation results: TPR, FPR, RMSE, RPE based on 100 replicates for balanced cases with $\sigma=0.5$. Each cell shows the mean (s.d.).}
\label{tab:ex2e5}
\resizebox{\textwidth}{!}{%
\begin{tabular}{cclllll}
\hline
                     & $n$                  & \multicolumn{1}{c}{Method} & \multicolumn{1}{c}{TPR} & \multicolumn{1}{c}{FPR} & \multicolumn{1}{c}{RMSE} & \multicolumn{1}{c}{RPE} \\ \hline
\multirow{12}{*}{S3} & \multirow{6}{*}{200} & Proposed                    & \textbf{1.000(0.000)}   & 0.001(0.001)            & \textbf{0.019(0.022)}    & \textbf{0.574(0.042)}   \\
                     &                      & L-MLR                      & 0.431(0.092)            & 0.030(0.012)            & 0.043(0.011)             & 2.350(0.523)            \\
                     &                      & MoE                        & 0.321(0.043)            & \textbf{0.000(0.000)}   & 0.057(0.018)             & 3.090(1.064)            \\
                     &                      & S-FMR                      & 0.380(0.142)            & 0.124(0.006)            & 0.073(0.002)             & 3.310(0.700)            \\
                     &                      & FCM1                       & 0.027(0.060)            & 0.001(0.002)            & 0.114(0.002)             & 4.892(0.241)            \\
                     &                      & FCM2                       & 0.045(0.044)            & 0.001(0.004)            & 0.110(0.003)             & 4.373(0.427)            \\ \cline{2-7} 
                     & \multirow{6}{*}{400} & Proposed                    & \textbf{1.000(0.000)}   & \textbf{0.000(0.000)}   & \textbf{0.012(0.013)}    & \textbf{0.535(0.022)}   \\
                     &                      & L-MLR                      & 0.446(0.072)            & 0.029(0.015)            & 0.035(0.034)             & 2.172(0.476)            \\
                     &                      & MoE                        & 0.593(0.110)            & 0.100(0.006)            & 0.038(0.032)             & 2.381(0.908)            \\
                     &                      & S-FMR                      & 0.452(0.145)            & 0.101(0.008)            & 0.066(0.049)             & 3.066(0.602)            \\
                     &                      & FCM1                       & 0.053(0.091)            & \textbf{0.000(0.000)}   & 0.111(0.001)             & 4.802(0.150)            \\
                     &                      & FCM2                       & 0.062(0.054)            & 0.001(0.001)            & 0.112(0.001)             & 4.473(0.203)            \\ \hline
\multirow{12}{*}{S4} & \multirow{6}{*}{200} & Proposed                    & \textbf{0.993(0.006)}   & \textbf{0.001(0.001)}   & \textbf{0.018(0.015)}    & \textbf{0.534(0.034)}   \\
                     &                      & L-MLR                      & 0.443(0.096)            & 0.021(0.011)            & 0.036(0.033)             & 2.525(0.680)            \\
                     &                      & MoE                        & 0.334(0.052)            & \textbf{0.001(0.001)}   & 0.056(0.011)             & 3.363(0.862)            \\
                     &                      & S-FMR                      & 0.373(0.105)            & 0.115(0.008)            & 0.069(0.004)             & 3.505(0.672)            \\
                     &                      & FCM1                       & 0.090(0.116)            & 0.001(0.003)            & 0.111(0.001)             & 4.541(0.215)            \\
                     &                      & FCM2                       & 0.122(0.171)            & 0.008(0.015)            & 0.109(0.007)             & 4.314(0.593)            \\ \cline{2-7} 
                     & \multirow{6}{*}{400} & Proposed                    & \textbf{1.000(0.000)}   & \textbf{0.000(0.000)}   & \textbf{0.013(0.002)}    & \textbf{0.520(0.020)}   \\
                     &                      & L-MLR                      & 0.450(0.095)            & 0.018(0.008)            & 0.035(0.149)             & 2.363(0.590)            \\
                     &                      & MoE                        & 0.667(0.106)            & 0.073(0.025)            & 0.040(0.013)             & 2.529(1.165)            \\
                     &                      & S-FMR                      & 0.544(0.118)            & 0.074(0.007)            & 0.055(0.074)             & 3.026(0.539)            \\
                     &                      & FCM1                       & 0.118(0.136)            & 0.032(0.002)            & 0.110(0.001)             & 4.510(0.183)            \\
                     &                      & FCM2                       & 0.157(0.185)            & 0.007(0.011)            & 0.106(0.009)             & 4.317(0.402)            \\ \hline
\end{tabular}%
}}
\end{table}

\noindent{\bf More examples}. \rm
To further test the effectiveness of the method for setting the initial grouping used in the proposed approach, we compare the performance of the proposed approach against two conventional techniques to mitigate the dependence on the initial grouping: (a) MIV (Multiple initial values), which draws multiple initial values randomly and chooses the one yielding the lowest objective function (\ref{eq:problem}), and (b) VNS (Variable neighborhood search)\cite[]{Bonhomme2015Grouped}, which adopts local searching and neighborhood jumping ideas. Table \ref{tab:initial} reports the simulation results under S1 setting. The results under S3 setting are reported in Table \ref{tab:initial2}. The numerical results suggest that the proposed approach is superior to the alternatives. The alternatives perform well when the number of features $p$ is small. However, their performance deteriorates significantly as $p$ increases. 

\begin{table}[h]
\centering
\caption{Simulation results of proposed approach, MIV with 50 randomly chosen starting value, and
VNS based on 100 replicates under S1 setting with $n = 200$ and $\sigma = 1$. Each cell
shows the mean (s.d.)}
\label{tab:initial}
\resizebox{\textwidth}{!}{%
\begin{tabular}{cccccccl}
\hline
$p$                   & Method   & $L_1$ loss            & TPR                   & FPR                   & RMSE                  & RPE                   & \multicolumn{1}{c}{Time(seconds)} \\ \hline
\multirow{3}{*}{10}   & Proposed & \textbf{0.150(0.032)} & \textbf{0.988(0.029)} & \textbf{0.031(0.021)} & \textbf{0.685(0.062)} & \textbf{1.683(0.196)} & \textbf{3.747(1.701)}             \\
                      & MIV      & 0.176(0.034)          & 0.954(0.039)          & 0.045(0.03)           & 0.689(0.066)          & 1.746(0.194)          & 18.288(2.810)                     \\
                      & VNS      & 0.166(0.033)          & 0.971(0.038)          & 0.039(0.022)          & 0.689(0.065)          & 1.752(0.187)          & 8.401(2.604)                      \\ \hline
\multirow{3}{*}{50}   & Proposed & \textbf{0.169(0.043)} & \textbf{0.963(0.031)} & \textbf{0.012(0.017)} & \textbf{0.334(0.067)} & \textbf{1.692(0.207)} & \textbf{14.968(4.673)}            \\
                      & MIV      & 0.206(0.05)           & 0.925(0.068)          & 0.021(0.028)          & 0.346(0.064)          & 1.788(0.342)          & 34.120(9.109)                     \\
                      & VNS      & 0.198(0.045)          & 0.936(0.062)          & 0.017(0.023)          & 0.35(0.064)           & 1.781(0.304)          & 30.641(8.528)                     \\ \hline
\multirow{3}{*}{100}  & Proposed & \textbf{0.171(0.044)} & \textbf{0.953(0.054)} & \textbf{0.011(0.009)} & \textbf{0.224(0.035)} & \textbf{1.703(0.339)} & \textbf{23.730(8.072)}            \\
                      & MIV      & 0.213(0.056)          & 0.826(0.104)          & 0.024(0.013)          & 0.234(0.054)          & 2.076(0.484)          & 285.351(85.099)                   \\
                      & VNS      & 0.203(0.054)          & 0.854(0.103)          & 0.022(0.011)          & 0.243(0.032)          & 2.070(0.407)          & 107.253(93.055)                   \\ \hline
\multirow{3}{*}{500}  & Proposed & \textbf{0.230(0.043)} & \textbf{0.905(0.100)} & \textbf{0.002(0.002)} & \textbf{0.102(0.012)} & \textbf{1.704(0.564)} & \textbf{106.956(13.462)}          \\
                      & MIV      & 0.267(0.056)          & 0.648(0.129)          & 0.004(0.003)          & 0.110(0.015)          & 2.092(0.596)          & 332.351(85.008)                   \\
                      & VNS      & 0.257(0.052)          & 0.669(0.128)          & 0.003(0.002)          & 0.110(0.014)          & 2.101(0.590)          & 215.868(106.559)                  \\ \hline
\multirow{3}{*}{1000} & Proposed & \textbf{0.263(0.052)} & \textbf{0.843(0.116)} & \textbf{0.001(0.002)} & \textbf{0.041(0.008)} & \textbf{1.701(0.648)} & \textbf{201.493(70.483)}          \\
                      & MIV      & 0.294(0.055)          & 0.416(0.132)          & 0.002(0.003)          & 0.084(0.009)          & 2.109(0.763)          & 571.413(137.082)                  \\
                      & VNS      & 0.288(0.058)          & 0.432(0.133)          & 0.002(0.002)          & 0.084(0.008)          & 2.109(0.694)          & 360.938(138.264)                  \\ \hline
\end{tabular}%
}
\end{table}

So far, what we study belong to the well-specified case in which the true model is exactly the proposed one, i.e., the outcomes are generated from the heterogeneous \emph{linear} regression models. To further test the robustness of the proposed approach, we also consider a misspecified case. Specifically, the $n=400$ samples are divided into two overlapping subgroups in the same way as Example 1. For the samples belonging to only one subgroup, their outcomes and covariates obey unique \emph{quadratic} regression models: 
\[
	y_i = x_{i1}^2\alpha_{k1} + \sum_{j=2}^p x_{ij}\alpha_{kj} + \epsilon, \text{ if }u_{ki} = 1.
\]
The true coefficient vectors are 
\[\bm A=\left(
\begin{array}{cc}
0.3 & 0.3 \\
1 & -1 \\
2 & -2 \\
0_{(p-3)\times 1} & 0_{(p-3)\times 1} \\
\end{array}
\right).\]
Figure \ref{fig:ms_kselect} shows the histograms of estimated number of subgroups $\hat{K}$. It is observed that the proposed approach tends to select the more complex model with a larger $\hat{K}$. {\color{blue} As pointed out in \cite{Nguyen2021b}, in the misspecified case, the best choice for $K$ should balance the trade-off between model bias and variance, thus usually leading to a more complex model (i.e., with more subgroups) as observed in our numerical results}.

Besides the heterogeneous models, we also examine the performance of {\color{blue}the} proposed approach on the homogeneous model, i.e., all data are generated from an identical linear regression model $y_i=X_i\alpha+\epsilon_i$, $i=1,\ldots,n$. The true coefficient matrix degenerates into a coefficient vector
\[\bm A=\left(
\begin{array}{c}
1 \\
2 \\
3 \\
0_{(p-3)\times 1}\\
\end{array}
\right).\]

For the homogeneous case, the true number of subgroups is 1. From Table \ref{tab:1k4kselect}, we observe that the median value of $\hat{K}$ is 1, and the mean value is close to 1, suggesting that the proposed approach is also suit for the homogeneous model.  

Finally, we apply the proposed approach to the low-dimensional data. Table \ref{tab:low_dim} presents the simulation results under S1 setting with $p=10,20,30,50,100$. Here the results of fusion method \cite[]{Ma2020} are provided for $p=10,20,30$; however, for larger $p$, methodological and computational difficulties are encountered, hence this method is omitted from our reporting. It can be seen that the proposed approach still has the best or close to the best performance.

\section{Real Data Analysis}
\subsection{Cancer Cell Line Encyclopedia data}
\label{sec:Cancer}
The Cancer Cell Line Encyclopedia (CCLE, Barretina\cite{Barretina2012}) contains 947 cancer cell lines from nine types of cancers with associated gene expression measurements and responses to 24 anti-cancer agents. We treat the area above the dose-response curve as the response and use the gene expression measurements as features. Here, we use these data to explore the underlying subgroups with respect to gene-response associations, and to identify the important genes for each subgroup. 

We apply the proposed approach to 24 anti-cancer drugs. After removing the cell lines with missing values, 400--500 samples are included for each drug. In the raw dataset, there are 18,899 gene expression measurements. Although the proposed approach could, in principle, be applied to all genes, the small sample size and additional complexity brought by heterogeneity may make the analysis unreliable. As such, we conduct a feature screening described in Section \ref{sec:initial}. Specifically, we select the top 120 sub-models based on the BIC criterion (\ref{eq:BIC1}), and we keep all the genes in these sub-models. A total of 600 genes are kept for the downstream analysis. 

The proposed approach identifies three to eight subgroups for each of the 24 drugs. Take response PF2341066 as an example, with the BIC criterion, four subgroups are identified by the proposed approach (Figure \ref{fig:bic_app} (a)). We show estimated weight matrix $\bm U$ in Figure \ref{fig:ccle_group_pf} (a). For most of samples, the weights are very close to 1 or 0, indicating that these cells can be assigned to a specific subgroup with low uncertainty. However, it can be observed that there indeed are some cells that have more prominent weights in at least two subgroups and thus cannot to be assigned to one subgroup (see also Figure \ref{fig:toy_example} (a)), which suggests the need for the proposed overlapping subgroup strategy. The estimated regression coefficient of the four subgroups are presented in Table \ref{tab:coef_CCLE}. Significant differences across the subgroups are observed. This result indicates the need to conduct heterogeneity analyses with respect to feature-response associations.

\begin{figure}[!tbp]
\setlength{\abovecaptionskip}{0.2cm}
\setlength{\belowcaptionskip}{0cm}
  \centering
  \includegraphics[width=0.8\textwidth]{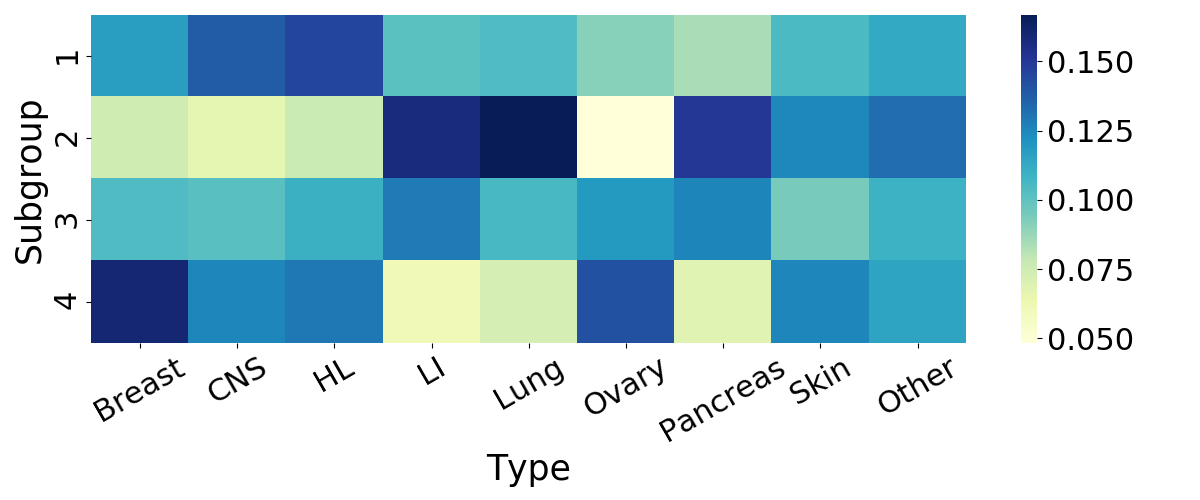}
  \caption{Analysis of CCLE data (response PF2341066) using the proposed approach: weighted proportion of cells with a certain cancer type in four subgroups (HL: haematopoietic and lymphoid tissue; CNS:central nervous system; LI: large intestine).}
  \label{fig:weight_group_pf}
\end{figure}

\begin{table}[!tbp]
\centering
\caption{Analysis of CCLE data (response PF2341066) using the proposed approach: identified genes and estimates for the four subgroups.}
\label{tab:coef_CCLE}
\begin{tabular}{lllll}
\hline
Gene         & Subgroup 1 & Subgroup 2 & Subgroup 3 & Subgroup 4 \\ \hline
HLA.A        & 0.414      &      &    &     \\
PIGM         & 0.234      &       &      &    \\
ATP6V1G2     & 0.130      &      &     &      \\
TRAPPC6A     & 0.104      &       &       &      \\
IDO1         & 0.092      &      &      & 0.052      \\
ITGA5        & -0.110     &       &       &      \\
FAM65A       & -0.106     &      &       &     \\
POMGNT1      & -0.103     & 1.050      &     &     \\
GOSR1        & -0.093     &      &       &     \\
RFX5         & -0.087     &       &       &       \\
C1orf200     & -0.083     &     &     &       \\
LOC100506302 &      & 0.659      &      &      \\
CYP2S1       &       & 0.286      &     &     \\
RFK          &    & 0.131      &       & 0.082      \\
FAM167B      &       & 0.123      &      & 0.080      \\
ATP8B1       &      & 0.080      &       &      \\
BANK1        &      & 0.056      &     & 0.578      \\
KRIT1        &    & -0.454     &     & -0.244     \\
FIBIN        &     & -0.271     &       &   \\
EFEMP2       &      & -0.169     &       &      \\
FAM65B       &       & -0.158     &      &      \\
PHYH         &      & -0.090     &      & 0.147      \\
GPR39        &      & -0.579     &      &      \\
PROS1        &     &       & 0.157      & 0.123      \\
DEDD         &       &      & 0.143      &       \\
SNHG8        &      &     & 0.099      &       \\
MTMR7        &     &       & 0.094      & 0.115      \\
CCNY         &      &       & 0.092      &       \\
TRPV4        &      &       & 0.073      &       \\
RASSF7       &     &      & 0.058      & 0.080      \\
GOLT1A       &      &      & -0.102     &    \\
SIPA1L2      &       &       & -0.066     &     \\
ANXA3        &      &       & -0.061     & -0.231     \\
MAGEA11      &       &       & -0.255     &      \\
FAM198B      &      &      &      & 0.274      \\
FTSJD1       &       &    &      & 0.254      \\
FRZB         &     &      &     & 0.223      \\
TAF4B        &       &    &      & 0.156      \\
GH1          &       &      &       & 0.134      \\
AGFG2        &    &     &      & 0.114      \\
UHRF2        &      &       &       & 0.112      \\
ZKSCAN2      &    &     &     & -0.160     \\
RRAGD        &     &       &      & -0.146     \\ \hline
\end{tabular}
\end{table}

To investigate whether the subgroups are biologically meaningful, we calculate the weighted proportion of cells in each subgroup that have the corresponding cancer type, and present the results in Figure \ref{fig:weight_group_pf}, where the proportion is weighted by the estimated $u_{ki}$. Differences across the four subgroups are clearly observed. Specifically, subgroups 1 and 2 have higher percentages of the central nervous system and haematopoietic and lymphoid tissue, and large intestine, lung and pancreas cancers, respectively; subgroup 4 is dominated by breast and ovary cancers, while subgroup 3 is distributed more evenly. 
Existing literature provides support to the validity of the results. For example, breast cancer and ovary cancer are known to share some gene mutations like BRCA1 and BRCA2, and thus are expected to react similarly to a certain anti-cancer agent. 

Besides the proposed approach, we also analyze data using the alternatives. {\color{blue} For S-FMR, the number of subgroups is set to two as is done in its original paper \cite[]{liu2020}. For other alternatives, the number of subgroups are determined by the same BIC-criterion (\ref{eq:bic}) over a grid of $K=1,2,\ldots, 6$}. The number of identified subgroups are 4(L-MLR), 3(MoE), 2(S-FMR), 4(FCM1), and 4(FCM2). Comparing the subgroup structures of the proposed and alternative approaches reveals significant differences (Figure \ref{fig:ccle_group_pf}). The feature selection results are compared in Table \ref{tab:compare}, where the significant distinction is again observed. As has been noted in published studies, it is challenging to objectively determine which set of results is more sensible. We conduct the following analysis, which may support the results to a large extent. We first evaluate the prediction performance. Specifically, we partition each dataset into a training and testing set with a size ratio of 7:3. Estimation is conducted using the training set, and prediction is made using the testing test. The process is repeated independently 100 times. Figure \ref{fig:rpe_box_all} presents the results for all 24 drugs. It can be observed that the proposed approach largely outperforms the alternatives in prediction. In addition, we examine the stability of subgroup identification. Specifically, 30\% of samples are randomly removed from each dataset. Let $\bm{G} = (g_{lm})_{n \times n}$ be the co-existence matrix, where the $(l,m)$th element $g_{lm}=1$ if samples $l,m$ belong to the same subgroup, and 0 otherwise. Here, each sample is assigned to the majority subgroup. We define the stability measure $\bm{M}_{sta} = 1/n^2\sum_{l,m=1}^n|\hat{\bm{G}}-\bm{G}^{(T)}|$, as suggested by Teran\cite{Teran2017}, where $\hat{\bm{G}}$ and $\bm{G}^{(T)}$ are the co-existence matrices of the estimated and true subgroups, respectively. The ``truth" is taken as the estimates using all samples. 
A smaller $\bm{M}_{sta}$ means higher stability. For response PF2341066, the mean values of $\bm{M}_{sta}$ over 100 random replicates are 0.194(Proposed), 0.289(L-FMR), 0.375(MoE), 0.356(S-FMR), 0.347(FCM1), and 0.358(FCM2). Similar patterns are observed for other responses. 

\subsection{The Cancer Genome Atlas lung cancer data}
\label{sec:Lung}
The Cancer Genome Atlas (TCGA) is a collaborative effort organized by NCI and has recently published high-quality profiling data on multiple cancer types. The heterogeneity analysis of the TCGA data has already been conducted \cite[]{Lawrence2013, Hofree2013Network}. However, the existing studies adopt clustering-based analysis, and there is a lack of regression-based heterogeneity analysis. Data analyzed here are downloaded from the cBioPortal website (www.cbioportal.org). 

We consider lung cancer data. Lung cancer is the second most common cancer in the world. As the proposed approach can accommodate data heterogeneity, 
we combine the Lung adenocarcinoma (LUAD) and Lung squamous cell carcinoma (LUSC) data, which are the most prevalent subtypes of lung cancer. 
It is noted that the two subtypes are defined mainly using pathological characteristics. In this analysis, we are interested in the heterogeneity in the regulation of FEV1 (forced expiratory volume in 1sec), a critical measure of lung function, by gene expressions. Existing studies do not suggest whether such heterogeneity is linked to pathology-based one or not. 

As in the literature, the inclusion criteria are (1) no neo-adjuvant therapy before tumor sample collection, (2) in stage I of the AJCC pathologic tumor stage measurement, and (3) with FEV1 and gene expressions measured. A total of 231 samples, and 20, 531 RNAseq gene expressions are available for analysis. We consider the processed level-3 gene expression data and refer to the literature \cite[]{Molony2009} for detailed information on data generation and processing. 
To improve the reliability of the results, we first conduct an ANOVA analysis to select differentially expressed tumor tissue genes compared with normal samples, which leads to 4230 gene expressions to the downstream analysis.

\begin{figure}[!tbp]
\setlength{\abovecaptionskip}{-0.2cm}
\setlength{\belowcaptionskip}{-0.cm}
  \centering
  \includegraphics[width=0.7\textwidth]{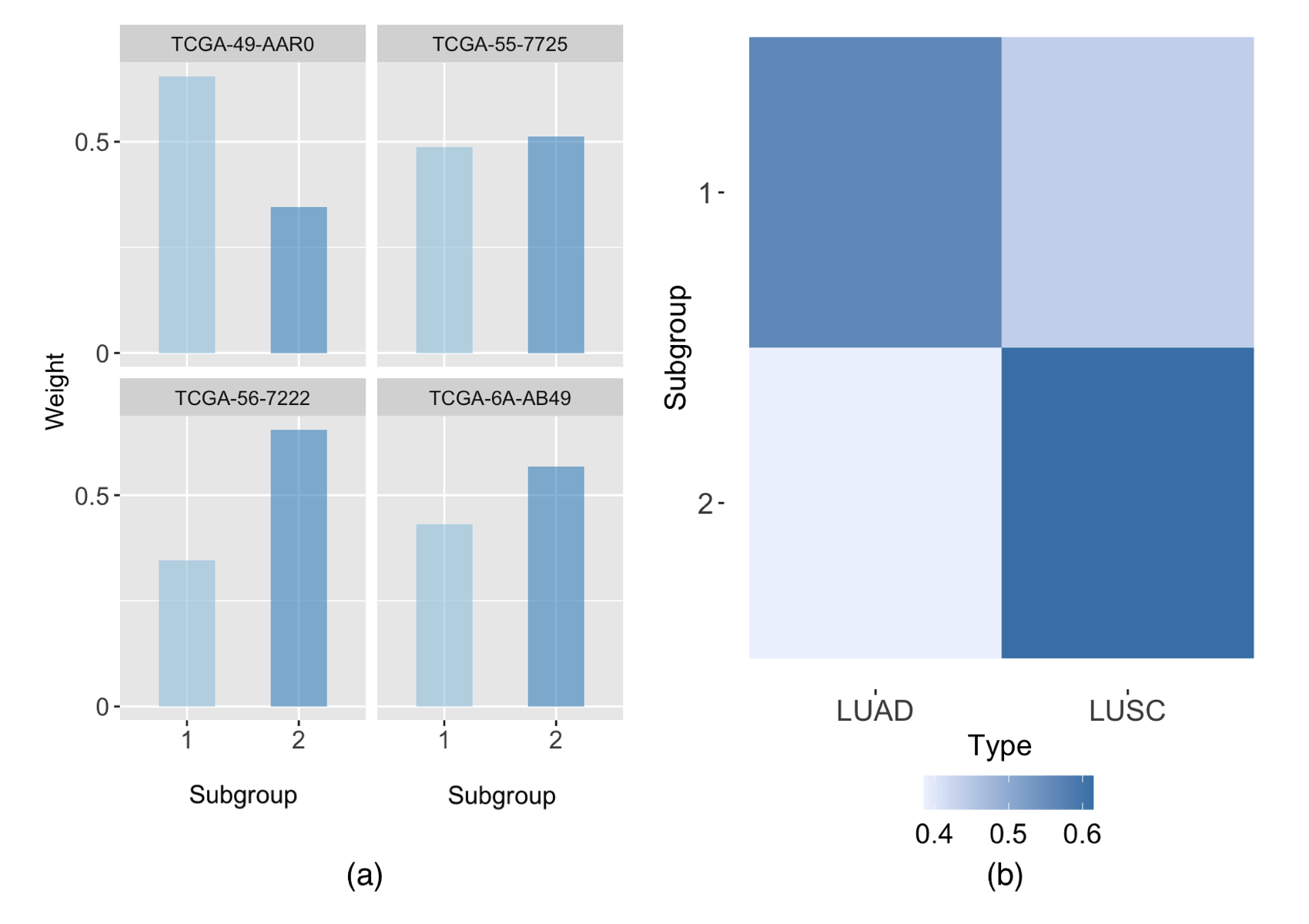}
  \caption{Analysis of TCGA lung cancer data using the proposed approach: (a) weights of four patients, and (b) weighted proportion of patients with a certain lung cancer subtype in the two subgroups.}
  \label{fig:rst_lung}
\end{figure}

With the BIC criterion, the proposed approach identifies two subgroups (Figure \ref{fig:bic_app} (b)). The identified important genes and corresponding estimates for the two subgroups are shown in Table \ref{tab:coef_lung}. Significant differences are observed again. Figure \ref{fig:lung_weights_heatmap} (a) presents the estimated weight matrix $\bm{U}$. It is noted that some patients have non-ignorable weights in both subgroups (Figure \ref{fig:rst_lung} (a)), which suggests again the necessity of heterogeneity analysis that accommodate overlapping-subgroups. 

To determine the biological importance of the estimated weights, we investigate whether they are predictive of observed clinical data. The identified subgroups are closely associated with the recorded subtypes LUAD/LUSC (Figure \ref{fig:rst_lung} (b); $\chi^2$ test, $P<10^{-8}$). The results, along with Figure \ref{fig:rst_lung} (a), indicate the existence of patients whose relationships between FEV1 and gene expression are mixed ones of LUAD and LUSC. We also fit a Cox-proportional hazards model to determine the relationship between the weights and patient survival. A likelihood-ratio test and associated P-value is calculated by comparing the full model, which includes individuals' weight vectors in subgroup 1 and clinical covariates including age, gender, race, age at index, and pack of cigarettes smoked per year, against a baseline model that includes covariates only. The likelihood ratio test ($P<0.01$) demonstrates that the estimated weights are predictive of survival independently of clinical covariates. 

Besides the proposed approach, we also analyze data using the alternatives. The number of identified subgroups are 3(L-MLR), 4(MoE), 2(S-FMR), 3(FCM1), and 3(FCM2). The differences in subgrouping results between the proposed and alternative approaches are observed (Figure \ref{fig:lung_weights_heatmap}). Only the subgroups identified by S-FMR are less significantly associated with the recorded subtypes ($\chi^2$ test, $P=0.01$), and others lead to an insignificant association. Besides, none of the subgroups identified by the alternatives is predictive of patient survival. The comparison of feature selection results in Table \ref{tab:compare} also suggests significant differences. In prediction performance, the average RPE values are 0.271(Proposed), 0.443(L-MLR), 0.591(MoE), 0.437(S-FMR), 0.927(FCM1), and 0.478(FCM2). Overall, it again seems that the proposed approach leads to more sensible findings. 

\subsection{Simulation based on CCLE data}
It has been recognized in some studies that simulated data may be ``simpler'' than real data. Here we conduct an additional set of simulation based on the CCLE data analyzed above. Specifically, we again take the response PF2341066 as an example. The observed cancer cell line measurements and estimated number of subgroups, obtained using the method proposed in Section \ref{sec:Cancer}, are used in the simulation. The sparsity structure of the coefficient matrix and weight matrix is the same as the ones estimated by the proposed method. The simulation results are summarized in Table \ref{tab:ccle_simu}. It is observed that the proposed method maintains a relative advantage over the alternatives, which demonstrates the effectiveness of the proposed method.

\section{Conclusion}
In this study, we have developed a novel regression-based approach for heterogeneity analysis. The proposed approach introduces a subgroup membership weight vector for each sample and combines it with a loss function to accommodate overlapping subgroups. A fusion-type penalization is developed to encourage weights to be close to each other. We have also applied additional penalization to accommodate the high data dimension and to screen out noises. Another contribution is the development of an effective method for setting initial values to tackle the initial value sensitivity. The simulation demonstrated its superiority over the direct competitors, and in the data analysis, it generates findings with improved prediction and stability in subgrouping. 

Although this study only presents the application of this method to genomics, the proposed analysis can also be conducted with other types of high-dimensional data, such as medical image data. The proposed approach can be potentially extended to other outcomes and models. 
The theoretical investigations of the proposed approach, e.g., identifiability and tuning parameter selection, are deferred to future research. In the real-data analysis, heterogeneity has been identified. For each dataset, several subgroups have been identified that differ in terms of the relationship between gene expressions and response. It is noted that some subgroups indeed overlap. The biological implications of these heterogeneous subgroups, especially the overlapping between subgroups, deserve further investigation.

\begin{acknowledgement}
This study was partly supported by the National Natural Science Foundation of China (12171479), Fund for building world-class universities (disciplines) of Renmin University of China (18XNB004). 
\end{acknowledgement}
\vspace*{1pc}

\noindent {\bf{Conflict of Interest}}

\noindent {\it{The authors have declared no conflict of interest.}}

\noindent {\bf{Data Availability}}

\noindent{The authors confirm that the data supporting the findings of this study are available within the article. }

\noindent {\it{}}

\section*{Appendix}
\setcounter{table}{0}
\renewcommand\thetable{A\arabic{table}}

\setcounter{figure}{0}
\renewcommand\thefigure{A\arabic{figure}}

\setcounter{algorithm}{0}
\renewcommand\thealgorithm{A\arabic{algorithm}}

\section*{A. More details on updating weight matrix\label{app2}}%

Given coefficient matrix $\bm A$, the optimization problem with respect to weight matrix $\bm U$ is as follows: 
\begin{equation}
\label{eq:problem_w}
\begin{aligned}
\min_{\bm U} & \sum_{k=1}^{K}\sum_{i=1}^{n}u_{ik}^m(y_i - X_i \alpha_k)^2 + \gamma \sum_{k=2}^K\sum_{i=1}^n  (u_{i,(k)}-u_{i,(k-1)})^2 \\
\text{s.t.}\ \ & \ \sum_{k=1}^K u_{ik} = 1, u_{ik} \geq 0\quad (i = 1,...,n; k = 1,...,K)\\
 \end{aligned}
 \end{equation}
 We use the interior-point method to solve the constrained optimization problem (\ref{eq:problem_w}). Denote 
 $U = (u_{11},u_{12},\ldots,$ \\
 $u_{1K},u_{21},u_{22},\ldots,u_{nK})^{\top}\in R^{nK\times 1}$. 
 The log-barrier function has the form 
 \[F(U)=\sum_{k=1}^{K}\sum_{i=1}^{n}u_{ik}^m(y_i - X_i \alpha_k)^2 + \gamma \sum_{k=2}^K\sum_{i=1}^n  (u_{i,(k)}-u_{i,(k-1)})^2-r\sum_{i=1}^n\sum_{k=1}^K \text{log}(u_{ik}),\]
where $r$ is the barrier parameter. Let $\mathbb{I}_k$ represent the $k\times 1$ vector with all elements being 1, and define $\bm M=\text{diag}(\underbrace{\mathbb{I}_k^{\top},...,\mathbb{I}_k^{\top}}_{n})$. Then, the solution to optimization problem (\ref{eq:problem_w}) can be obtained by solving the following equality-constrained optimization problem as $r\to 0$ \cite[]{Boyd2009}: 

\begin{equation}
\label{eq:problem_app}
\begin{aligned}
\min_{U} \ \ & F(U)  \\
\text{s.t.}\ \ & \  \bm{M}U = E_n\\
 \end{aligned}
 \end{equation}

We use the Newton method to solve (\ref{eq:problem_app}). The algorithm is summarized in Algorithm \ref{alg3}, where $\triangle U = U^*{(r)} - U^{(k)}$, $\nabla F(U^{(k)})$ and $\nabla^2 F(U^{(k)})$ are the first and second derivative of $F(U^{(k)})$, respectively.

\begin{algorithm}[H]
\caption{Interior-point method to solve the optimization problem (\ref{eq:problem_w})}\label{alg3}
0. Initialization. Set $k=0, r=1$, and tolerance $\epsilon > 0$. Initialize $U^{(k)}$. 

Repeat

\ \ 1. Compute $U^*(r)$ with the standard Newton method starting at $U^{(k)}$:

\ \ Repeat

\qquad\ \ 1.1 Compute the Newton step $\delta U_{k}$: 
$\delta U_{k} = -\nabla^2F(U^{(k)})^{-1} (\nabla F(U^{(k)})+\bm M^{\top}\Gamma)$, where $ \Gamma = (\bm M\nabla^2F(U^{(k)})^{-1}\bm M)^{-1}(-\bm M\nabla^2F(U^{(k)})^{-1}\nabla F(U^{(k)})^{-1})$. 

\qquad\ \ 1.2 Compute the decrement $\lambda(U) = \delta U_{k}^{\top}\|\nabla^2 F(U^{(k)})\|^2_2  \delta U_{k}$.

\qquad\ \ 1.3 Quit if $\lambda^2/2 \leq \epsilon$.

\qquad\ \ 1.4 Choose step size $t$ by backtracking line search:

\qquad\qquad 1.4.1 Set $t = 1$.

\qquad\qquad 1.4.2 Update $U^*(r) \leftarrow U^{(k)} + t \delta U_k$.

\qquad\qquad 1.4.3 Increase step size by $t \leftarrow 0.8t$ until
\begin{equation*}
F(U^*(r))  \leq F(U^{(k)}) + \nabla F(U^{(k)}) \triangle U + \frac{1}{2t}\|\triangle U\|^2_2 	
\end{equation*}

\qquad\ \ 1.5 Update. $U^{(k)} \leftarrow U^*(r)$.

\ \ 2. Set $k = k + 1$. $U^{(k)}\leftarrow U^{(k-1)}$.

\ \ 3. Quit if $nKr < \epsilon$.

\ \ 4. Update $r \leftarrow  r/10$.
\end{algorithm}

\section*{B. Variable neighborhood search\label{app3}}%

The variable neighborhood search (VNS) algorithm was first proposed to solve the minimum sum-of-squares partitioning problem \cite[]{2001Variable}. Recently, it has been applied to low-dimensional panel data with a grouped structure. Here, we extend the specific algorithm used in \cite[]{Bonhomme2015Grouped} to high-dimensional data with grouped structure. The algorithm works as follows: 

\begin{algorithm}
\caption{VNS}\label{alg_app}
1. Initialization. Perform Algorithm A1 with a random initial value, and let the obtained parameter values be the starting values $({\bm U}^{0},\bm{A}^{0})$. Set $iter_{max}=5$ and $neigh_{max}=15$, and $iter=0$.

2. Set $neigh=1$.

3. Relocate $n$ randomly selected samples to $n$ randomly selected groups and obtain new weight matrix ${\bm U}^{'}$. Perform Step 2 of Algorithm \ref{alg2} and obtain new coefficient matrix $\bm{A}^{'}$.

4. Apply Algorithm \ref{alg2} using $({\bm U}^{'},\bm{A}^{'})$ as starting values, and obtain new parameter values $({\bm U}^{''},\bm{A}^{''})$.

5. If $\text{BIC}({\bm U}^{''},\bm{A}^{''} )<{\text{BIC}}({\bm U}^{0},\bm{A}^{0})$, then set $({\bm U}^{0}, {\bm A}^{0})=({\bm U}^{''},\bm{A}^{''})$ and go back to Step 3; otherwise set $neigh=neigh+1$ and go to Step 6.

6. If $neigh \leq neigh_{max},$ then go to Step 3; otherwise, go to Step 7. 

7. Set $iter=iter+1$. If $iter>iter_{max}$, then stop; otherwise go back to Step 2.
\end{algorithm}

\section*{C. More tables and figures} 
\label{appC}

\begin{table}[h]
{\tiny
\centering
\caption{Simulation results: TPR, FPR, RMSE, RPE based on 100 replicates with $\sigma=1$. Each cell shows the mean (s.d.).}
\label{tab:ex1e1}
\resizebox{\textwidth}{!}{%
\begin{tabular}{cclllll}
\hline
                     & $n$                  & \multicolumn{1}{c}{Method} & \multicolumn{1}{c}{TPR} & \multicolumn{1}{c}{FPR} & \multicolumn{1}{c}{RMSE} & \multicolumn{1}{c}{RPE} \\ \hline
\multirow{12}{*}{S1} & \multirow{6}{*}{200} & Propsed                    & \textbf{0.843(0.116)}   & \textbf{0.001(0.002)}   & \textbf{0.041(0.008)}    & \textbf{1.701(0.648)}   \\
                     &                      & L-MLR                      & 0.376(0.092)            & 0.148(0.061)            & 0.082(0.018)             & 3.143(0.581)            \\
                     &                      & MoE                        & 0.308(0.039)            & 0.003(0.001)            & 0.102(0.019)             & 4.793(0.498)            \\
                     &                      & S-FMR                      & 0.101(0.084)            & 0.120(0.072)            & 0.131(0.013)             & 3.436(0.902)            \\
                     &                      & FCM1                       & 0.027(0.061)            & 0.006(0.008)            & 0.201(0.006)             & 7.995(0.515)            \\
                     &                      & FCM2                       & 0.048(0.089)            & 0.013(0.013)            & 0.118(0.008)             & 7.145(0.425)            \\ \cline{2-7} 
                     & \multirow{6}{*}{400} & Propsed                    & \textbf{0.917(0.050)}   & 0.001(0.001)            & \textbf{0.034(0.006)}    & \textbf{1.435(0.427)}   \\
                     &                      & L-MLR                      & 0.385(0.082)            & 0.137(0.075)            & 0.074(0.022)             & 2.938(0.553)            \\
                     &                      & MoE                        & 0.394(0.025)            & \textbf{0.000(0.000)}   & 0.071(0.011)             & 4.208(0.347)            \\
                     &                      & S-FMR                      & 0.117(0.101)            & 0.092(0.051)            & 0.120(0.029)             & 3.095(0.942)            \\
                     &                      & FCM1                       & 0.033(0.076)            & 0.006(0.010)            & 0.189(0.007)             & 7.901(0.410)            \\
                     &                      & FCM2                       & 0.087(0.071)            & 0.012(0.013)            & 0.103(0.011)             & 7.359(0.419)            \\ \hline
\multirow{12}{*}{S2} & \multirow{6}{*}{200} & Propsed                    & \textbf{0.775(0.206)}   & 0.003(0.009)            & \textbf{0.054(0.026)}    & \textbf{1.810(0.499)}   \\
                     &                      & L-MLR                      & 0.358(0.133)            & 0.158(0.081)            & 0.092(0.021)             & 2.322(0.643)            \\
                     &                      & MoE                        & 0.244(0.047)            & 0.005(0.002)            & 0.116(0.017)             & 4.188(0.648)            \\
                     &                      & S-FMR                      & 0.102(0.006)            & 0.282(0.095)            & 0.133(0.004)             & 2.841(0.135)            \\
                     &                      & FCM1                       & 0.020(0.054)            & \textbf{0.001(0.002)}   & 0.119(0.003)             & 4.271(0.214)            \\
                     &                      & FCM2                       & 0.047(0.075)            & 0.002(0.005)            & 0.120(0.007)             & 4.18(0.395)             \\ \cline{2-7} 
                     & \multirow{6}{*}{400} & Propsed                    & \textbf{0.863(0.153)}   & 0.004(0.016)            & \textbf{0.044(0.210)}    & \textbf{1.569(0.212)}   \\
                     &                      & L-MLR                      & 0.366(0.102)            & 0.143(0.057)            & 0.073(0.012)             & 2.059(0.630)            \\
                     &                      & MoE                        & 0.396(0.048)            & 0.008(0.010)            & 0.088(0.012)             & 3.151(1.894)            \\
                     &                      & S-FMR                      & 0.123(0.006)            & 0.230(0.104)            & 0.125(0.008)             & 2.298(0.073)            \\
                     &                      & FCM1                       & 0.027(0.070)            & \textbf{0.001(0.001)}   & 0.082(0.002)             & 4.285(0.171)            \\
                     &                      & FCM2                       & 0.077(0.096)            & \textbf{0.001(0.006)}   & 0.086(0.006)             & 4.216(0.349)            \\ \hline
\end{tabular}%
}}
\end{table}

\begin{table}[h]
\centering
\caption{ Simulation results: TPR, FPR, RMSE, RPE, and $L_1$ loss based on 100 replicates under overlapping subgroups setting with small coefficients,  $n=400$, and $\sigma = 0.5$. Each cell shows the mean (s.d.)}
\label{tab:low_beta1}
\resizebox{\textwidth}{!}{%
\begin{tabular}{lccccc}
\hline
          \multicolumn{1}{c}{Method} & TPR                   & FPR                   & RMSE                  & RPE                   & $L_1$ loss            \\ \hline
 Proposed                   & \textbf{0.906(0.092)} & 0.003(0.003)          & \textbf{0.016(0.010)} & \textbf{0.548(0.114)} & \textbf{0.397(0.127)} \\
                      L-MLR                      & 0.418(0.123)          & 0.113(0.009)          & 0.024(0.017)          & 1.957(0.362)          & 0.457(0.210)          \\
                      MoE                        & 0.296(0.214)          & 0.094(0.088)          & 0.028(0.022)          & 2.031(1.448)          & 0.488(0.241)          \\
                      S-FMR                      & 0.143(0.076)          & 0.082(0.013)          & 0.059(0.014)          & 2.113(0.160)          & 0.871(0.279)          \\
                      FCM1                       & 0.033(0.073)          & \textbf{0.002(0.005)} & 0.083(0.002)          & 3.376(0.205)          & 0.949(0.040)          \\
                      FCM2                       & 0.108(0.101)          & 0.013(0.013)          & 0.076(0.004)          & 2.849(0.166)          & 0.973(0.020)          \\ \hline

\end{tabular}%
}
\end{table}

\begin{table}[h]
\centering
\caption{ Simulation results: the sample mean, median, and standard deviation (s.d.) of $\hat{K}$ and the percentage of $\hat{K}$ equaling the true number of subgroups by the proposed approach based on 100 replicates with $\sigma = 0.5$.}
\label{tab:1k4kselect}
\begin{tabular}{@{}ccccc@{}}
\toprule
K & Mean & Median & s.d. & Percentage \\ \midrule
1 & 1.35 & 1.00   & 0.85 & 0.76       \\
4 & 4.45 & 4.00   & 1.09 & 0.62       \\ \bottomrule
\end{tabular}
\end{table}

\begin{table}[h]
\centering
\caption{ Simulation results: TPR, FPR, RMSE, RPE, and $L_1$ loss based on 100 replicates under overlapping subgroups setting with $K=4$, $n=900$ and $\sigma = 0.5$. Each cell shows the mean (s.d.)}
\label{tab:4k}
\begin{tabular}{lccccc}
\hline
         & TPR                   & FPR                   & RMSE                  & RPE                   & $L_1$ loss            \\ \hline
Proposed & \textbf{0.847(0.047)} & 0.056(0.007)          & \textbf{0.122(0.006)} & \textbf{1.085(0.216)} & \textbf{0.769(0.045)} \\
L-MLR    & 0.387(0.098)          & 0.109(0.075)          & 0.156(0.008)          & 3.469(0.479)          & 0.954(0.109)          \\
MoE      & 0.267(0.057)          & \textbf{0.002(0.003)} & 0.163(0.006)          & 4.249(0.559)          & 1.053(0.095)          \\
S-FMR    & 0.333(0.059)          & 0.045(0.008)          & 0.153(0.003)          & 3.655(0.264)          & 0.950(0.094)          \\
FCM1     & 0.001(0.081)          & 0.007(0.005)          & 0.206(0.005)          & 5.331(0.262)          & 1.509(0.081)          \\
FCM2     & 0.079(0.027)          & 0.016(0.010)          & 0.190(0.003)          & 5.235(0.319)          & 1.489(0.113)          \\ \hline
\end{tabular}
\end{table}

\begin{table}[h]
{\tiny
\centering
\caption{Simulation results: TPR, FPR, RMSE, RPE based on 100 replicates for balanced cases with $\sigma=1$. Each cell shows the mean (s.d.)}
\label{tab:ex2e1}
\resizebox{\textwidth}{!}{%
\begin{tabular}{cclllll}
\hline
                     & $n$                  & \multicolumn{1}{c}{Method} & \multicolumn{1}{c}{TPR} & \multicolumn{1}{c}{FPR} & \multicolumn{1}{c}{RMSE} & \multicolumn{1}{c}{RPE} \\ \hline
\multirow{12}{*}{S3} & \multirow{6}{*}{200} & Propsed                    & \textbf{0.955(0.030)}   & 0.002(0.002)            & \textbf{0.030(0.021)}    & \textbf{1.095(0.095)}   \\
                     &                      & L-MLR                      & 0.362(0.091)            & 0.034(0.017)            & 0.079(0.014)             & 3.655(1.215)            \\
                     &                      & MoE                        & 0.317(0.042)            & \textbf{0.001(0.001)}   & 0.062(0.016)             & 3.371(1.064)            \\
                     &                      & S-FMR                      & 0.357(0.130)            & 0.127(0.017)            & 0.087(0.013)             & 3.713(0.953)            \\
                     &                      & FCM1                       & 0.027(0.078)            & \textbf{0.001(0.002)}   & 0.118(0.003)             & 4.959(0.195)            \\
                     &                      & FCM2                       & 0.045(0.045)            & 0.002(0.005)            & 0.117(0.005)             & 4.411(0.434)            \\ \cline{2-7} 
                     & \multirow{6}{*}{400} & Propsed                    & \textbf{1.000(0.000)}   & \textbf{0.000(0.000)}   & \textbf{0.018(0.003)}    & \textbf{1.045(0.045)}   \\
                     &                      & L-MLR                      & 0.360(0.115)            & 0.029(0.012)            & 0.071(0.026)             & 2.539(0.973)            \\
                     &                      & MoE                        & 0.400(0.096)            & 0.002(0.001)            & 0.041(0.059)             & 2.729(2.290)            \\
                     &                      & S-FMR                      & 0.413(0.128)            & 0.099(0.007)            & 0.072(0.008)             & 3.558(0.813)            \\
                     &                      & FCM1                       & 0.037(0.080)            & \textbf{0.000(0.000)}   & 0.115(0.004)             & 4.875(0.164)            \\
                     &                      & FCM2                       & 0.052(0.049)            & 0.001(0.001)            & 0.113(0.007)             & 4.536(0.215)            \\ \hline
\multirow{12}{*}{S4} & \multirow{6}{*}{200} & Propsed                    & \textbf{1.000(0.000)}   & 0.002(0.000)            & \textbf{0.018(0.015)}    & \textbf{1.034(0.079)}   \\
                     &                      & L-MLR                      & 0.322(0.124)            & 0.009(0.007)            & 0.074(0.024)             & 3.429(0.811)            \\
                     &                      & MoE                        & 0.333(0.032)            & 0.002(0.001)            & 0.069(0.011)             & 3.392(0.274)            \\
                     &                      & S-FMR                      & 0.363(0.121)            & 0.114(0.008)            & 0.086(0.014)             & 3.996(0.812)            \\
                     &                      & FCM1                       & 0.105(0.133)            & \textbf{0.001(0.003)}   & 0.112(0.003)             & 4.504(0.257)            \\
                     &                      & FCM2                       & 0.095(0.174)            & 0.006(0.012)            & 0.111(0.007)             & 4.443(0.535)            \\ \cline{2-7} 
                     & \multirow{6}{*}{400} & Propsed                    & \textbf{1.000(0.000)}   & \textbf{0.000(0.000)}   & \textbf{0.011(0.002)}    & \textbf{1.006(0.034)}   \\
                     &                      & L-MLR                      & 0.350(0.132)            & 0.008(0.006)            & 0.071(0.023)             & 2.827(0.741)            \\
                     &                      & MoE                        & 0.483(0.098)            & 0.003(0.002)            & 0.044(0.031)             & 2.741(1.051)            \\
                     &                      & S-FMR                      & 0.447(0.103)            & 0.076(0.016)            & 0.070(0.011)             & 3.285(0.866)            \\
                     &                      & FCM1                       & 0.112(0.142)            & \textbf{0.000(0.000)}   & 0.109(0.005)             & 4.460(0.184)            \\
                     &                      & FCM2                       & 0.127(0.173)            & 0.008(0.012)            & 0.109(0.012)             & 4.386(0.402)            \\ \hline
\end{tabular}%
}}
\end{table}

\begin{table}[h]
{\tiny
\centering
\caption{Simulation results: TPR, FPR, RMSE, RPE based on 100 replicates for unbalanced cases with $\sigma=0.5$. Each cell shows the mean (s.d.).}
\label{tab:ex2e5pr7}
\resizebox{\textwidth}{!}{%
\begin{tabular}{cclllll}
\hline
                     & $n$                  & \multicolumn{1}{c}{Method} & \multicolumn{1}{c}{TPR} & \multicolumn{1}{c}{FPR} & \multicolumn{1}{c}{RMSE} & \multicolumn{1}{c}{RPE} \\ \hline
\multirow{12}{*}{S3} & \multirow{6}{*}{200} & Proopsed                    & \textbf{1.000(0.000)}   & 0.001(0.001)            & \textbf{0.022(0.003)}    & \textbf{0.560(0.039)}   \\
                     &                      & L-MLR                      & 0.440(0.064)            & 0.029(0.020)            & 0.035(0.006)             & 2.280(0.300)            \\
                     &                      & MoE                        & 0.343(0.041)            & 0.003(0.002)            & 0.059(0.003)             & 3.092(0.215)            \\
                     &                      & S-FMR                      & 0.410(0.118)            & 0.118(0.008)            & 0.076(0.005)             & 3.349(0.761)            \\
                     &                      & FCM1                       & 0.023(0.057)            & \textbf{0.000(0.000)}   & 0.116(0.003)             & 4.859(0.211)            \\
                     &                      & FCM2                       & 0.048(0.086)            & 0.004(0.008)            & 0.114(0.003)             & 4.005(0.524)            \\ \cline{2-7} 
                     & \multirow{6}{*}{400} & Proposed                    & \textbf{1.000(0.000)}   & \textbf{0.000(0.000)}   & \textbf{0.014(0.003)}    & \textbf{0.530(0.019)}   \\
                     &                      & L-MLR                      & 0.443(0.053)            & 0.029(0.013)            & 0.034(0.012)             & 2.226(0.907)            \\
                     &                      & MoE                        & 0.467(0.094)            & 0.010(0.004)            & 0.040(0.045)             & 2.817(1.125)            \\
                     &                      & S-FMR                      & 0.466(0.092)            & 0.076(0.013)            & 0.054(0.009)             & 3.064(0.817)            \\
                     &                      & FCM1                       & 0.035(0.008)            & 0.019(0.005)            & 0.115(0.004)             & 4.850(0.135)            \\
                     &                      & FCM2                       & 0.051(0.074)            & 0.001(0.003)            & 0.113(0.003)             & 4.277(0.291)            \\ \hline
\multirow{12}{*}{S4} & \multirow{6}{*}{200} & Proposed                    & \textbf{0.998(0.001)}   & \textbf{0.000(0.000)}   & \textbf{0.015(0.003)}    & \textbf{0.569(0.226)}   \\
                     &                      & L-MLR                      & 0.465(0.133)            & 0.008(0.007)            & 0.039(0.013)             & 2.499(0.625)            \\
                     &                      & MoE                        & 0.320(0.043)            & 0.005(0.003)            & 0.054(0.012)             & 3.295(1.002)            \\
                     &                      & S-FMR                      & 0.324(0.135)            & 0.107(0.007)            & 0.050(0.004)             & 3.318(0.744)            \\
                     &                      & FCM1                       & 0.103(0.133)            & 0.013(0.002)            & 0.113(0.002)             & 4.565(0.260)            \\
                     &                      & FCM2                       & 0.137(0.158)            & 0.006(0.006)            & 0.110(0.010)             & 3.192(0.634)            \\ \cline{2-7} 
                     & \multirow{6}{*}{400} & Proposed                    & \textbf{1.000(0.000)}   & \textbf{0.000(0.000)}   & \textbf{0.013(0.011)}    & \textbf{0.521(0.024)}   \\
                     &                      & L-MLR                      & 0.458(0.118)            & 0.008(0.008)            & 0.033(0.011)             & 2.431(0.831)            \\
                     &                      & MoE                        & 0.500(0.059)            & 0.006(0.005)            & 0.039(0.008)             & 2.420(1.523)            \\
                     &                      & S-FMR                      & 0.497(0.068)            & 0.075(0.010)            & 0.038(0.009)             & 3.067(0.723)            \\
                     &                      & FCM1                       & 0.108(0.122)            & 0.022(0.006)            & 0.111(0.001)             & 4.471(0.189)            \\
                     &                      & FCM2                       & 0.158(0.115)            & 0.003(0.005)            & 0.108(0.003)             & 3.070(0.430)            \\ \hline
\end{tabular}%
}}
\end{table}

\newpage

\begin{table}[h]
{\tiny
\centering
\caption{Simulation results: TPR, FPR, RMSE, RPE based on 100 replicates for unbalanced cases with $\sigma=1$. Each cell shows the mean (s.d.).}
\label{tab:ex2e1pr7}
\resizebox{\textwidth}{!}{%
\begin{tabular}{cclllll}
\hline
                     & $n$                  & \multicolumn{1}{c}{Method} & \multicolumn{1}{c}{TPR} & \multicolumn{1}{c}{FPR} & \multicolumn{1}{c}{RMSE} & \multicolumn{1}{c}{RPE} \\ \hline
\multirow{12}{*}{S3} & \multirow{6}{*}{200} & Proposed                    & \textbf{0.973(0.028)}   & 0.003(0.002)            & \textbf{0.025(0.005)}    & \textbf{1.086(0.096)}   \\
                     &                      & L-MLR                      & 0.369(0.095)            & 0.033(0.011)            & 0.072(0.009)             & 2.985(0.608)            \\
                     &                      & MoE                        & 0.313(0.044)            & \textbf{0.000(0.000)}   & 0.067(0.028)             & 3.359(1.133)            \\
                     &                      & S-FMR                      & 0.343(0.114)            & 0.120(0.007)            & 0.087(0.006)             & 3.781(0.684)            \\
                     &                      & FCM1                       & 0.017(0.033)            & 0.002(0.015)            & 0.118(0.001)             & 4.875(0.217)            \\
                     &                      & FCM2                       & 0.045(0.065)            & 0.001(0.002)            & 0.117(0.002)             & 4.258(0.395)            \\ \cline{2-7} 
                     & \multirow{6}{*}{400} & Proposed                    & \textbf{1.000(0.000)}   & \textbf{0.001(0.002)}   & \textbf{0.017(0.002)}    & \textbf{1.026(0.053)}   \\
                     &                      & L-MLR                      & 0.366(0.086)            & 0.031(0.009)            & 0.066(0.011)             & 2.567(0.804)            \\
                     &                      & MoE                        & 0.417(0.079)            & 0.005(0.003)            & 0.043(0.029)             & 3.533(1.818)            \\
                     &                      & S-FMR                      & 0.447(0.117)            & 0.084(0.013)            & 0.080(0.008)             & 3.345(0.899)            \\
                     &                      & FCM1                       & 0.028(0.067)            & 0.042(0.013)            & 0.117(0.005)             & 4.810(0.168)            \\
                     &                      & FCM2                       & 0.051(0.080)            & \textbf{0.001(0.003)}   & 0.115(0.002)             & 4.350(0.304)            \\ \hline
\multirow{12}{*}{S4} & \multirow{6}{*}{200} & Proposed                    & \textbf{0.995(0.002)}   & 0.005(0.004)            & \textbf{0.023(0.013)}    & \textbf{1.060(0.255)}   \\
                     &                      & L-MLR                      & 0.338(0.124)            & 0.009(0.005)            & 0.075(0.010)             & 2.999(0.812)            \\
                     &                      & MoE                        & 0.300(0.059)            & \textbf{0.000(0.000)}   & 0.066(0.019)             & 3.512(1.550)            \\
                     &                      & S-FMR                      & 0.321(0.131)            & 0.110(0.012)            & 0.082(0.006)             & 3.703(0.763)            \\
                     &                      & FCM1                       & 0.102(0.140)            & 0.017(0.003)            & 0.113(0.003)             & 4.563(0.258)            \\
                     &                      & FCM2                       & 0.085(0.132)            & 0.003(0.004)            & 0.107(0.005)             & 3.405(0.622)            \\ \cline{2-7} 
                     & \multirow{6}{*}{400} & Proposed                    & \textbf{1.000(0.000)}   & \textbf{0.000(0.000)}   & \textbf{0.013(0.009)}    & \textbf{1.027(0.254)}   \\
                     &                      & L-MLR                      & 0.343(0.091)            & 0.008(0.006)            & 0.073(0.022)             & 2.330(1.069)            \\
                     &                      & MoE                        & 0.417(0.113)            & 0.010(0.009)            & 0.042(0.059)             & 2.932(1.467)            \\
                     &                      & S-FMR                      & 0.430(0.114)            & 0.089(0.014)            & 0.078(0.006)             & 3.515(0.649)            \\
                     &                      & FCM1                       & 0.113(0.118)            & 0.005(0.006)            & 0.110(0.001)             & 4.438(0.164)            \\
                     &                      & FCM2                       & 0.123(0.139)            & 0.002(0.003)            & 0.101(0.003)             & 3.232(0.460)            \\ \hline
\end{tabular}%
}}
\end{table}

\begin{table}[h]
\centering
\caption{ Simulation results: TPR, FPR, RMSE, RPE, and ARI based on 100 replicates under disjoint subgroups (balanced structure) with small coefficients,  $n=400$, and $\sigma = 0.5$. Each cell shows the mean (s.d.)}
\label{tab:low_beta2}
\resizebox{\textwidth}{!}{%
\begin{tabular}{lccccc}
\hline

 \multicolumn{1}{c}{Method} & TPR                   & FPR                   & RMSE                  & RPE                   & ARI                   \\ \hline
Proposed                   & \textbf{0.936(0.097)} & 0.003(0.006)          & \textbf{0.010(0.010)} & \textbf{0.521(0.080)} & \textbf{0.679(0.090)} \\
L-MLR                      & 0.459(0.061)          & 0.056(0.004)          & 0.020(0.008)          & 0.978(0.181)          & 0.156(0.100)          \\
MoE                        & 0.314(0.217)          & 0.084(0.108)          & 0.023(0.022)          & 2.031(1.948)          & 0.215(0.199)          \\
S-FMR                      & 0.335(0.076)          & 0.072(0.013)          & 0.049(0.014)          & 1.813(0.160)          & 0.109(0.017)          \\
FCM1                       & 0.110(0.139)          & \textbf{0.001(0.002)} & 0.085(0.003)          & 3.588(0.137)          & -0.001(0.002)         \\
FCM2                       & 0.153(0.324)          & 0.008(0.015)          & 0.078(0.005)          & 3.209(0.344)          & 0.002(0.001)          \\ \hline
\end{tabular}%
}
\end{table}

\begin{table}[h]
\centering
\caption{Simulation results of proposed approach, MIV with 50 randomly chosen starting value, and VNS based on 100 {\color{blue}replicates} under S3 setting (balanced structure) with $n = 200$ and $\sigma = 1$. Each cell shows the mean (s.d.).}
\label{tab:initial2}
\resizebox{\textwidth}{!}{%
\begin{tabular}{clllllll}
\hline
\textbf{$p$}          & \multicolumn{1}{c}{Method} & \multicolumn{1}{c}{ARI} & \multicolumn{1}{c}{TPR} & \multicolumn{1}{c}{FPR} & \multicolumn{1}{c}{RMSE} & \multicolumn{1}{c}{RPE} & \multicolumn{1}{c}{Time(seconds)} \\ \hline
\multirow{3}{*}{10}   & Proposed                            & \textbf{0.741(0.038)}            & \textbf{1.000(0.000)}            & \textbf{0.070(0.043)}            & \textbf{0.301(0.050)}             & \textbf{1.021(0.035)}            & \textbf{2.093(1.304)}                      \\
                      & MIV                                 & 0.736(0.036)                     & \textbf{1.000(0.000)}            & 0.140(0.062)                     & 0.315(0.043)                      & 1.095(0.037)                     & 16.978(3.938)                              \\
                      & VNS                                 & 0.732(0.036)                     & \textbf{1.000(0.000)}            & 0.140(0.062)                     & 0.341(0.044)                      & 1.103(0.036)                     & 6.661(2.459)                               \\ \hline
\multirow{3}{*}{50}   & Proposed                            & \textbf{0.726(0.044)}            & \textbf{1.000(0.000)}            & \textbf{0.043(0.026)}            & \textbf{0.093(0.062)}             & \textbf{0.963(0.039)}            & \textbf{13.622(4.473)}                     \\
                      & MIV                                 & 0.711(0.043)                     & \textbf{1.000(0.000)}            & 0.064(0.036)                     & 0.116(0.043)                      & 0.971(0.042)                     & 30.996(7.66)                               \\
                      & VNS                                 & 0.708(0.046)                     & \textbf{1.000(0.000)}            & 0.064(0.035)                     & 0.121(0.034)                      & 0.977(0.046)                     & 32.778(8.16)                               \\ \hline
\multirow{3}{*}{100}  & Proposed                            & \textbf{0.717(0.056)}            & \textbf{1.000(0.000)}            & \textbf{0.031(0.022)}            & \textbf{0.070(0.016)}             & \textbf{0.978(0.032)}            & \textbf{24.332(7.391)}                     \\
                      & MIV                                 & 0.652(0.057)                     & 0.883(0.052)                     & 0.036(0.026)                     & 0.095(0.017)                      & 1.094(0.305)                     & 279.958(86.952)                            \\
                      & VNS                                 & 0.673(0.058)                     & 0.916(0.065)                     & 0.036(0.025)                     & 0.098(0.015)                      & 1.017(0.283)                     & 105.992(92.201)                            \\ \hline
\multirow{3}{*}{500}  & Proposed                            & \textbf{0.704(0.051)}            & \textbf{0.995(0.037)}            & \textbf{0.006(0.006)}            & \textbf{0.043(0.011)}             & \textbf{0.991(0.041)}            & \textbf{109.041(14.306)}                   \\
                      & MIV                                 & 0.503(0.284)                     & 0.692(0.124)                     & 0.012(0.008)                     & 0.049(0.027)                      & 1.704(0.939)                     & 342.87(88.404)                             \\
                      & VNS                                 & 0.512(0.322)                     & 0.791(0.121)                     & 0.012(0.007)                     & 0.047(0.026)                      & 1.375(0.508)                     & 207.99(107.51)                             \\ \hline
\multirow{3}{*}{1000} & Proposed                            & \textbf{0.697(0.069)}            & \textbf{0.955(0.030)}            & \textbf{0.002(0.002)}            & \textbf{0.030(0.021)}             & \textbf{1.095(0.095)}            & \textbf{200.768(74.939)}                   \\
                      & MIV                                 & 0.393(0.356)                     & 0.600(0.173)                     & 0.009(0.010)                     & 0.047(0.020)                      & 2.257(1.494)                     & 527.597(121.762)                           \\
                      & VNS                                 & 0.405(0.366)                     & 0.633(0.185)                     & 0.010(0.007)                     & 0.044(0.024)                      & 1.956(0.810)                     & 366.784(127.31)                            \\ \hline
\end{tabular}%
}
\end{table}

\begin{table}[h]
{\tiny
\centering
\caption{ Simulation results:  TPR, FPR, RMSE, RPE, and $L_1$ loss based on 100 {\color{blue}replicates} under S1 setting with $n = 200$ and $\sigma = 1$. Each cell shows the mean (s.d.).}
\label{tab:low_dim}
\resizebox{\textwidth}{!}{%
\begin{tabular}{cllllll}
\hline
p                    & \multicolumn{1}{c}{Method} & \multicolumn{1}{c}{TPR}        & \multicolumn{1}{c}{FPR}        & \multicolumn{1}{c}{RMSE}       & \multicolumn{1}{c}{RPE} & \multicolumn{1}{c}{$L_1$ loss} \\ \hline
\multirow{7}{*}{10}  & Proposed                   & 0.988(0.029)                   & \textbf{0.031(0.021)}          & \textbf{0.685(0.062)}          & 1.683(0.196)            & \textbf{0.150(0.032)}          \\
                     & Penalized Fusion           & \textbf{1.000(0.000)}          & 1.000(0.000)                   & 0.838(0.060)                   & \textbf{1.377(0.189)}   & 0.340(0.037)                   \\
                     & L-MLR                      & 0.964(0.052)                   & 0.100(0.087)                   & 0.715(0.110)                   & 1.894(0.211)            & 0.328(0.034)                   \\
                     & MoE                        & 0.903(0.065)                   & 0.121(0.095)                   & 0.747(0.102)                   & 2.053(0.482)            & 0.333(0.054)                   \\
                     & S-FMR                      & 0.930(0.069)                   & 0.169(0.164)                   & 0.732(0.104)                   & 1.915(0.329)            & 0.364(0.031)                   \\
                     & FCM1                       & 0.002(0.001)                   & 0.175(0.134)                   & 1.860(0.144)                   & 7.746(0.457)            & 0.921(0.069)                   \\
                     & FCM2                       & 0.003(0.001)                   & 0.242(0.143)                   & 1.585(0.364)                   & 6.812(0.829)            & 0.938(0.044)                   \\ \hline
\multirow{7}{*}{20}  & Proposed                   & 0.983(0.030)                   & \textbf{0.028(0.018)}          & \textbf{0.587(0.063)}          & 1.688(0.197)            & \textbf{0.158(0.037)}          \\
                     & Penalized Fusion           & \textbf{1.000(0.000)}          & 1.000(0.000)                   & 0.677(0.058)                   & \textbf{1.075(0.231)}   & 0.401(0.041)                   \\
                     & L-MLR                      & 0.933(0.064)                   & 0.102(0.088)                   & 0.601(0.068)                   & 1.837(0.201)            & 0.346(0.049)                   \\
                     & MoE                        & 0.897(0.086)                   & 0.140(0.101)                   & 0.690(0.108)                   & 2.713(0.391)            & 0.391(0.056)                   \\
                     & S-FMR                      & 0.901(0.080)                   & 0.173(0.067)                   & 0.621(0.106)                   & 1.869(0.264)            & \textbf{0.375(0.031)}          \\
                     & FCM1                       & 0.004(0.002)                   & 0.108(0.169)                   & 1.285(0.076)                   & 7.767(0.645)            & 0.915(0.043)                   \\
                     & FCM2                       & 0.005(0.003)                   & 0.150(0.123)                   & 1.189(0.128)                   & 6.988(0.95)             & 0.94(0.042)                    \\ \hline
\multirow{7}{*}{30}  & Proposed                   & 0.975(0.031)                   & \textbf{0.022(0.018)}          & \textbf{0.486(0.065)}          & 1.690(0.202)            & \textbf{0.162(0.040)}          \\
                     & Penalized Fusion           & \textbf{1.000(0.000)}          & 1.000(0.000)                   & 0.546(0.059)                   & \textbf{1.632(0.224)}   & 0.371(0.044)                   \\
                     & L-MLR                      & 0.907(0.065)                   & 0.104(0.069)                   & 0.508(0.076)                   & 2.034(0.687)            & 0.439(0.067)                   \\
                     & MoE                        & 0.833(0.127)                   & 0.134(0.088)                   & 0.543(0.083)                   & 2.846(0.444)            & 0.444(0.072)                   \\
                     & S-FMR                      & 0.885(0.076)                   & 0.190(0.078)                   & 0.519(0.100)                   & 2.070(0.747)            & 0.401(0.085)                   \\
                     & FCM1                       & 0.003(0.001)                   & 0.058(0.095)                   & 1.070(0.059)                   & 7.807(0.590)            & 0.933(0.052)                   \\
                     & FCM2                       & 0.005(0.003)                   & 0.067(0.093)                   & 0.919(0.160)                   & 7.140(0.911)            & 0.937(0.052)                   \\ \hline
\multirow{7}{*}{50}  & Proposed                   & \textbf{0.963(0.031)}          & \textbf{0.012(0.017)}          & \textbf{0.334(0.067)}          & \textbf{1.692(0.207)}   & \textbf{0.169(0.043)}          \\
                     & Penalized Fusion           & \multicolumn{1}{c}{\textbf{-}} & \multicolumn{1}{c}{\textbf{-}} & \multicolumn{1}{c}{\textbf{-}} & \multicolumn{1}{c}{-}   & \multicolumn{1}{c}{-}          \\
                     & L-MLR                      & 0.881(0.074)                   & 0.099(0.086)                   & 0.380(0.079)                   & 1.874(0.488)            & 0.384(0.064)                   \\
                     & MoE                        & 0.813(0.109)                   & 0.121(0.093)                   & 0.400(0.074)                   & 2.919(0.738)            & 0.488(0.072)                   \\
                     & S-FMR                      & 0.849(0.130)                   & 0.198(0.053)                   & 0.398(0.105)                   & 1.983(0.725)            & 0.409(0.093)                   \\
                     & FCM1                       & 0.004(0.004)                   & 0.058(0.095)                   & 0.848(0.047)                   & 7.916(0.571)            & 0.909(0.075)                   \\
                     & FCM2                       & 0.007(0.005)                   & 0.075(0.077)                   & 0.817(0.178)                   & 6.910(0.949)            & 0.943(0.035)                   \\ \hline
\multirow{7}{*}{100} & Proposed                   & \textbf{0.953(0.054)}          & \textbf{0.011(0.009)}          & \textbf{0.224(0.035)}          & \textbf{1.703(0.339)}   & \textbf{0.171(0.044)}          \\
                     & Penalized Fusion           & \multicolumn{1}{c}{-}          & \multicolumn{1}{c}{-}          & \multicolumn{1}{c}{-}          & \multicolumn{1}{c}{-}   & \multicolumn{1}{c}{-}          \\
                     & L-MLR                      & 0.856(0.105)                   & 0.097(0.085)                   & 0.332(0.078)                   & 1.863(0.646)            & 0.391(0.072)                   \\
                     & MoE                        & 0.733(0.112)                   & 0.091(0.078)                   & 0.397(0.065)                   & 3.092(0.860)            & 0.525(0.078)                   \\
                     & S-FMR                      & 0.741(0.110)                   & 0.169(0.063)                   & 0.340(0.089)                   & 2.029(0.664)            & 0.454(0.098)                   \\
                     & FCM1                       & 0.006(0.008)                   & 0.067(0.062)                   & 0.584(0.033)                   & 7.738(0.582)            & 0.940(0.041)                   \\
                     & FCM2                       & 0.010(0.007)                   & 0.054(0.079)                   & 0.575(0.174)                   & 6.538(1.113)            & 0.937(0.051)                   \\ \hline
\end{tabular}%
}}
\end{table}

\begin{table}[h]
\centering
\caption{ Analysis of CCLE data (response PF2341066) and TCGA lung cancer data: comparison of feature selection results. Each cell shows the number of overlapping identifications.}
\label{tab:compare}
\begin{tabular}{@{}lllllll@{}}
\hline
         & \multicolumn{1}{c}{Proposed} & \multicolumn{1}{c}{L-MLR} & \multicolumn{1}{c}{MoE} & \multicolumn{1}{c}{S-FMR} & \multicolumn{1}{c}{FCM1} & \multicolumn{1}{c}{FCM2} \\ \hline
\multicolumn{7}{c}{CCLE data}\\
 Proposed & 43                           & 15                        & 8                       & 13                        & 6                        & 11                       \\
L-MLR    &                              & 34                        & 7                       & 13                        & 5                        & 11                       \\
MoE      &                              &                           & 40                      & 22                        & 4                        & 6                        \\
S-FMR    &                              &                           &                         & 172                       & 11                       & 29                       \\
FCM1     &                              &                           &                         &                           & 23                       & 10                       \\
FCM2     &                              &                           &                         &                           &                          & 76                       \\ \hline
\multicolumn{7}{c}{TCGA lung cancer data}\\
Proposed & 29                           & 8                         & 6                       & 9                         & 2                        & 12                       \\
L-MLR    &                              & 53                        & 11                      & 8                         & 4                        & 11                       \\
MoE      &                              &                           & 43                      & 7                         & 5                        & 8                        \\
S-FMR    &                              &                           &                         & 28                        & 3                        & 5                        \\
FCM1     &                              &                           &                         &                           & 13                       & 4                        \\
FCM2     &                              &                           &                         &                           &                          & 62                       \\ \hline
\end{tabular}
\end{table}

\begin{table}[h]
\centering
\caption{Analysis of TCGA lung cancer data using the proposed approach: identified genes and estimates for the two subgroups.}
\label{tab:coef_lung}
\begin{tabular}{lll}
\hline
Gene       & Subgroup 1 & Subgroup 2 \\ \hline
ALPL       & 0.329      & 0.298      \\
SCGB3A1    & 0.126      &       \\
SBSN       & 0.101      &      \\
GOLPH3L    & 0.040      &       \\
QRICH1     & 0.034      &      \\
EGFL6      & -0.310     & -0.221     \\
DRG1       & -0.147     &      \\
UNC13D     & -0.143     &      \\
TP53AIP1   & -0.103     & -0.039     \\
LHFPL3-AS2 & -0.079     &      \\
MASP1      & -0.070     &      \\
MAGEA3     & -0.065     &      \\
COLGALT2   & -0.051     &       \\
EDN2       & -0.050     &      \\
SCG5       & -0.048     &      \\
EHHADH     & -0.032     &       \\
ALDH1A1    &       & 0.315      \\
ACKR1      &      & 0.127      \\
FCGR3B     &       & 0.051      \\
LINC00472  &       & 0.042      \\
IL33       &      & 0.022      \\
FKBP4      &      & -0.262     \\
CSAG3      &      & -0.096     \\
POLE3      &      & -0.082     \\
ARL6IP6    &       & -0.080     \\
GPD2       &       & -0.062     \\
SPC25      &      & -0.031     \\
RHNO1      &       & -0.023     \\
HMOX1      &       & -0.014     \\ \hline
\end{tabular}
\end{table}

\begin{table}[h]
\centering
\caption{Simulation based on the CCLE data (response PF2341066). Each cell shows the mean (s.d.).}
\label{tab:ccle_simu}
\begin{tabular}{@{}lccccc@{}}
\toprule
\multicolumn{1}{c}{Method} & TPR                   & FPR                   & RMSE                  & RPE                   & $L_1$ loss               \\ \midrule
Proposed                   & \textbf{0.707(0.069)} & 0.032(0.013)          & \textbf{0.069(0.008)} & \textbf{0.904(0.052)} & \textbf{1.109(0.063)} \\
L-MLR                      & 0.342(0.118)          & 0.135(0.054)          & 0.078(0.006)          & 1.909(0.076)          & 1.461(0.078)          \\
MoE                        & 0.312(0.057)          & 0.117(0.034)          & 0.086(0.005)          & 2.324(0.048)          & 1.488(0.058)          \\
S-FMR                      & 0.279(0.027)          & 0.021(0.002)          & 0.075(0.001)          & 1.941(0.031)          & 1.471(0.022)          \\
FCM1                       & 0.081(0.102)          & \textbf{0.017(0.038)} & 0.077(0.002)          & 1.848(0.017)          & 1.551(0.010)          \\
FCM2                       & 0.251(0.155)          & 0.042(0.032)          & 0.076(0.002)          & 1.688(0.131)          & 1.468(0.020)          \\ \bottomrule
\end{tabular}
\end{table}

\newpage
\begin{figure}[h]
\setlength{\belowcaptionskip}{-0.cm}
  \centering
  \includegraphics[width=0.6\textwidth]{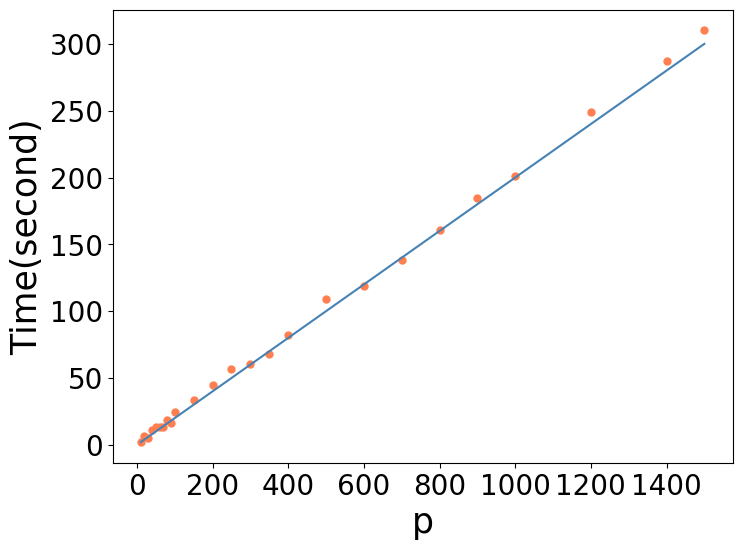}
  \caption{Simulation results: computation time of the proposed approach as a function of the number of features $p$ for one replicate under S4 setting with $n=200$ and $\sigma=0.5$.}
  \label{fig:time}
\end{figure}

\begin{figure}[h]
\setlength{\abovecaptionskip}{-0.2cm}
\setlength{\belowcaptionskip}{-0.cm}
  \centering
  \includegraphics[scale = 0.4]{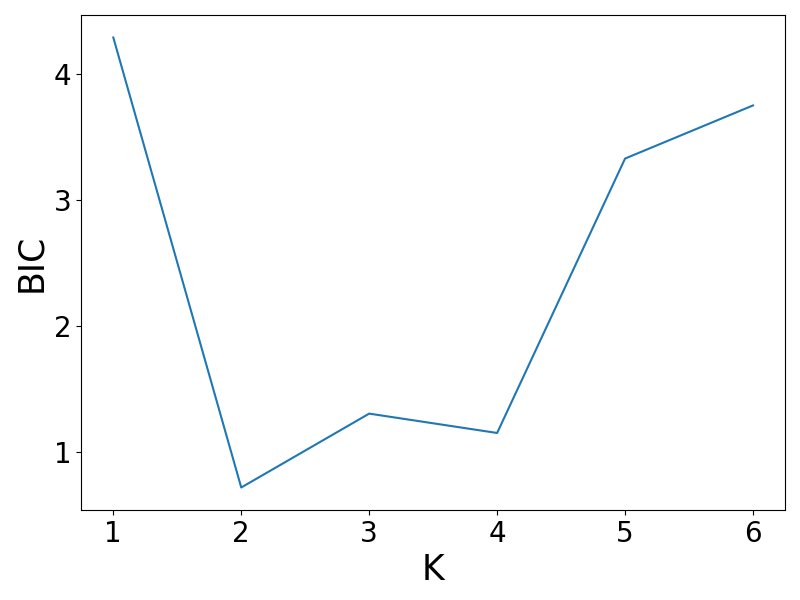}
  \caption{Simulation results: BIC as a function of the number of subgroups $K$ for one replicate under S1 setting with $n=200$ and $\sigma=0.5$.}
  \label{bic_simu}
\end{figure}

\begin{figure}[htp]
\setlength{\belowcaptionskip}{-0.cm}
  \centering
  \includegraphics[width=0.9\textwidth]{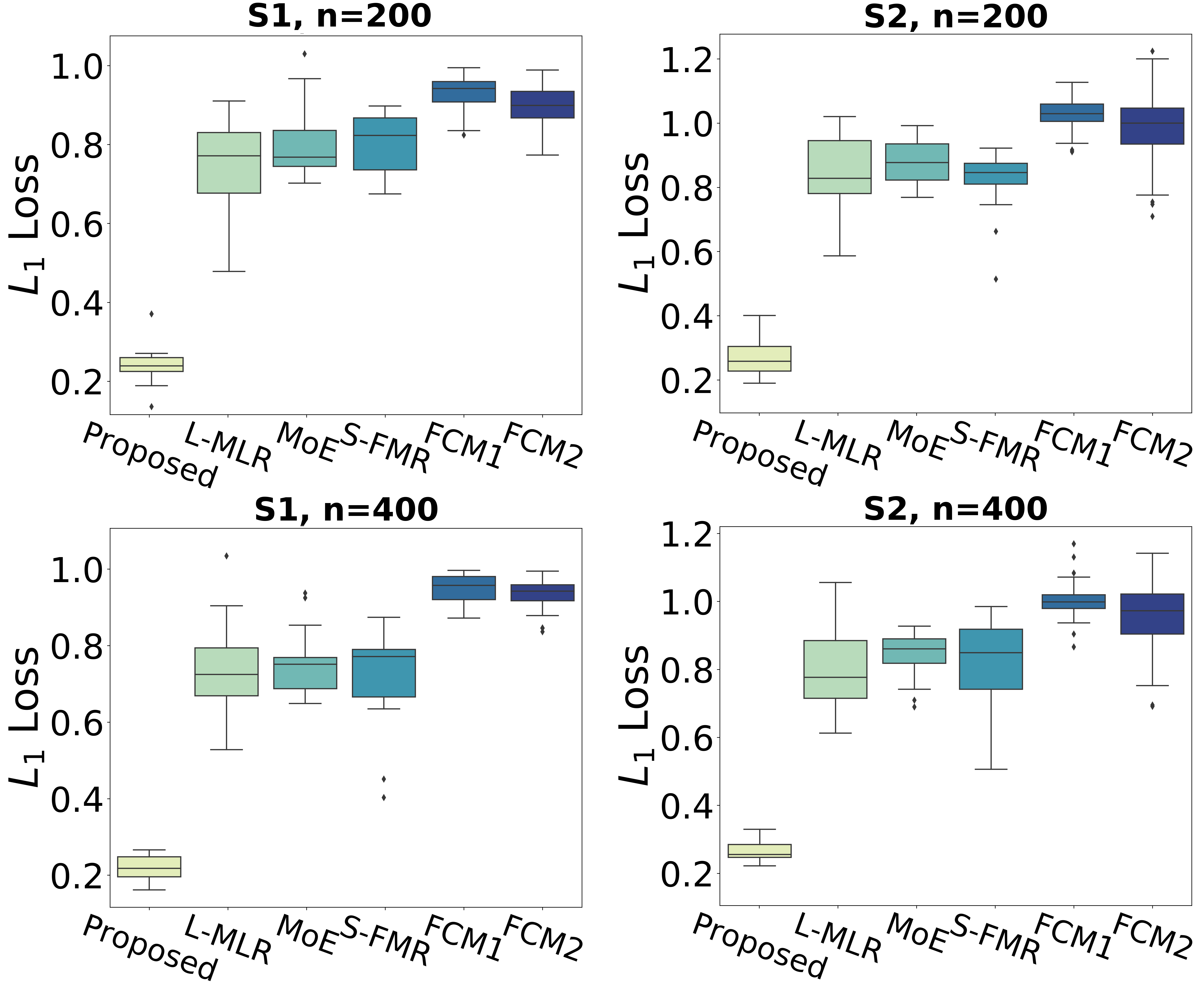}
  \caption{Simulation results: $L_1$ loss of estimating weight matrix $\bm U$ based on 100 {\color{blue}replicates} with $\sigma=0.5$.}
  \label{fig:l1loss_ex1e5}
\end{figure}

\begin{figure}[htp]
\setlength{\belowcaptionskip}{-0.cm}
  \centering
  \includegraphics[width=0.9\textwidth]{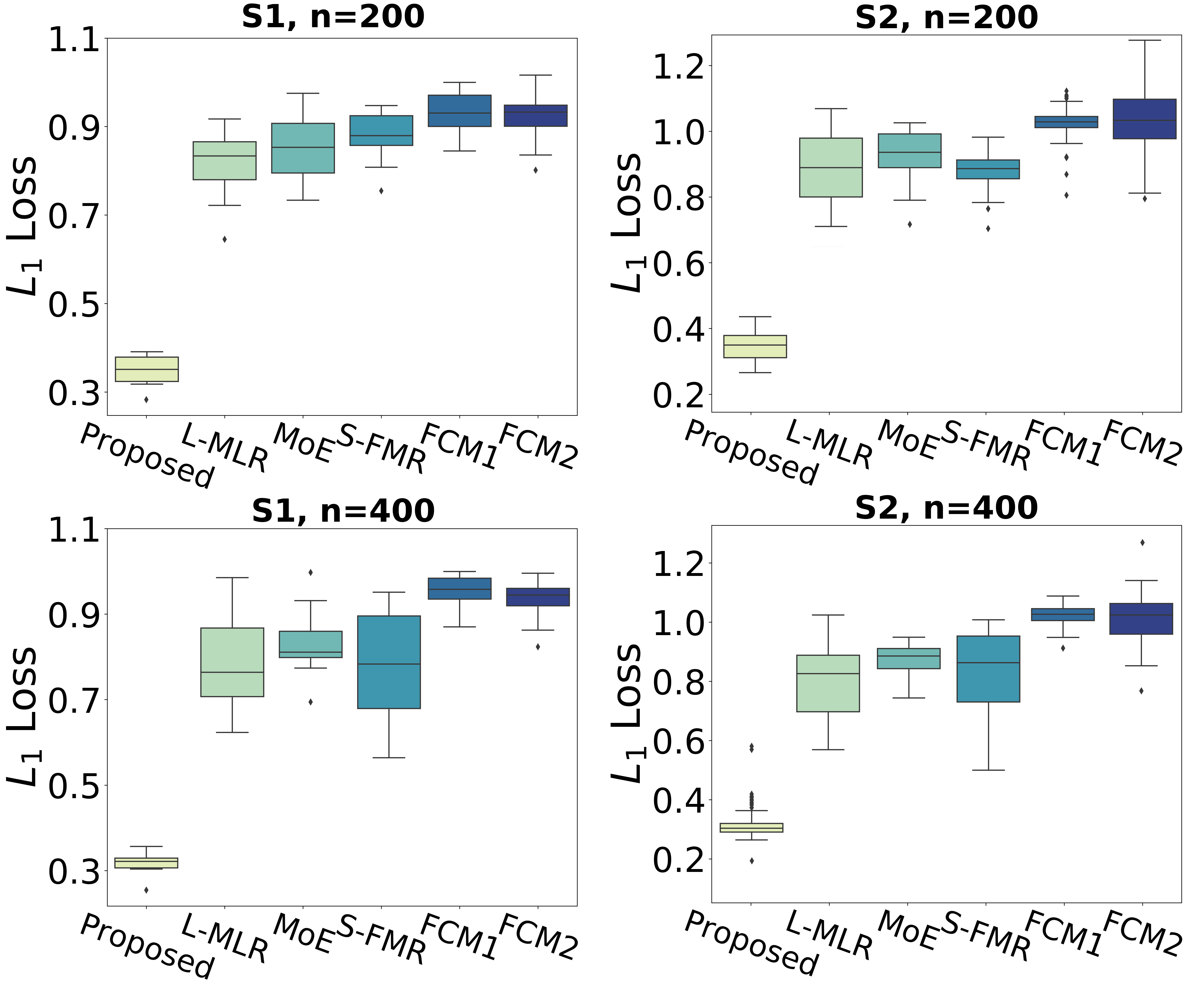}
  \caption{Simulation results: $L_1$ loss of estimating weight matrix $\bm U$ based on 100 replicates with $\sigma=1$.}
  \label{fig:l1loss_ex1e1}
\end{figure}

\begin{figure}[htp]
\setlength{\belowcaptionskip}{-0.cm}
  \centering
  \includegraphics[width=0.6\textwidth]{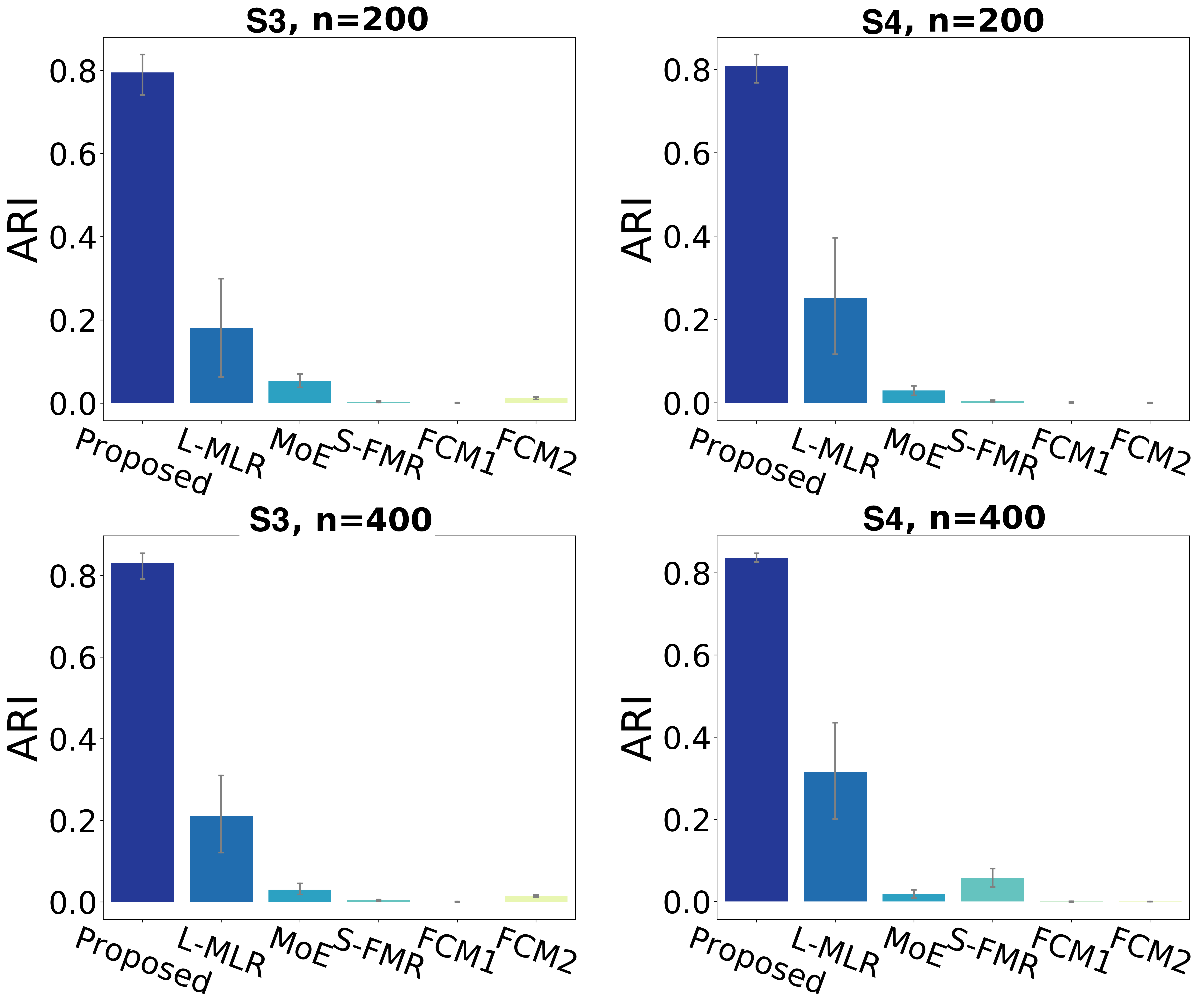}
   \caption{Simulation results: ARI based on 100 replicates for balanced cases with $\sigma=0.5$.}
  \label{fig:ari_ex2e5pr5}
\end{figure}

\begin{figure}[htp]
\setlength{\belowcaptionskip}{-0.cm}
  \centering
  \includegraphics[width=0.6\textwidth]{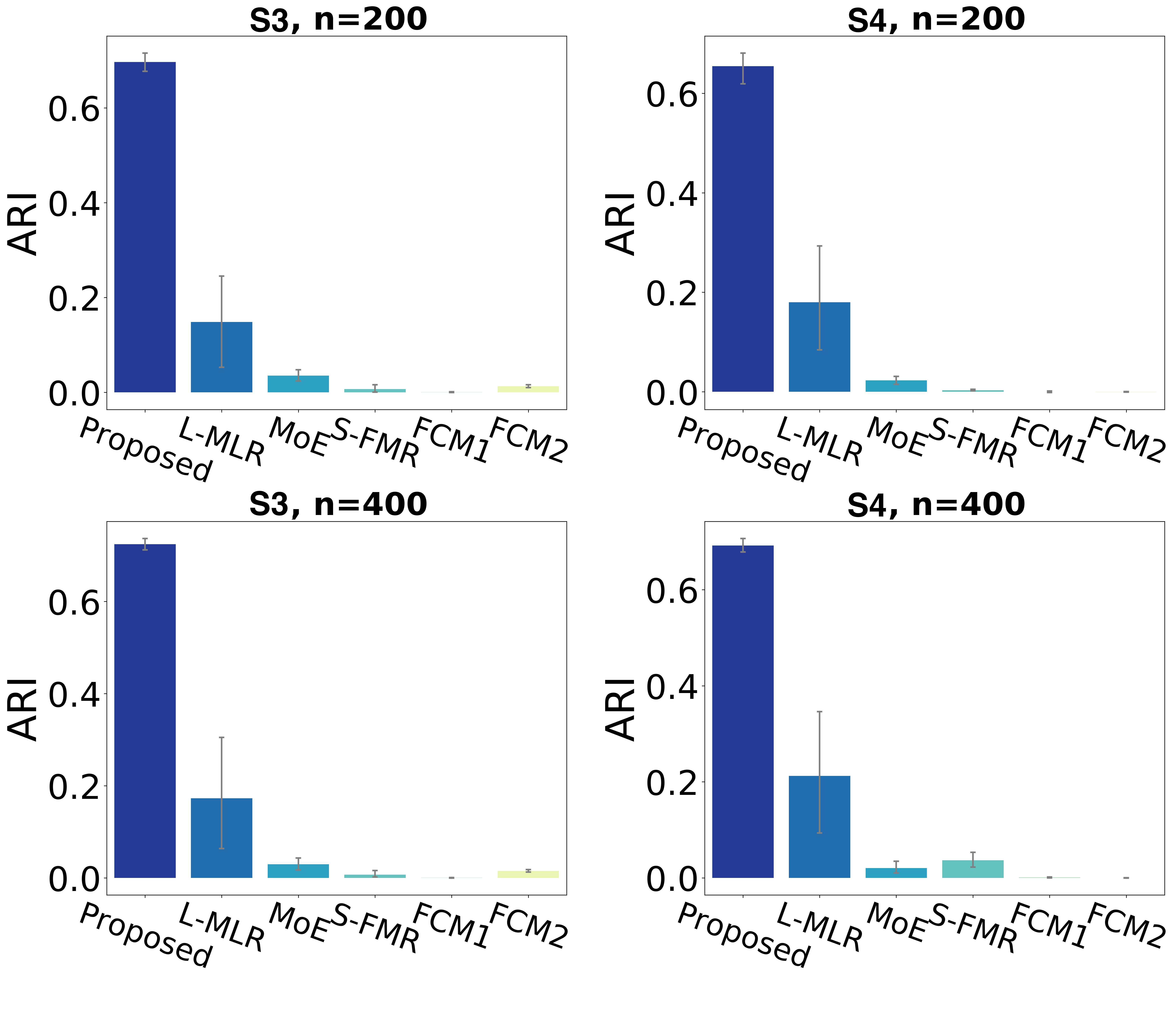}
  \caption{Simulation results: ARI based on 100 replicates for balanced cases with $\sigma=1$.}
  \label{fig:ari_ex2e1pr5}
\end{figure}

\newpage
\begin{figure}[htp]
\setlength{\abovecaptionskip}{-0.2cm}
\setlength{\belowcaptionskip}{-0.cm}
  \centering
  \includegraphics[scale = 0.5]{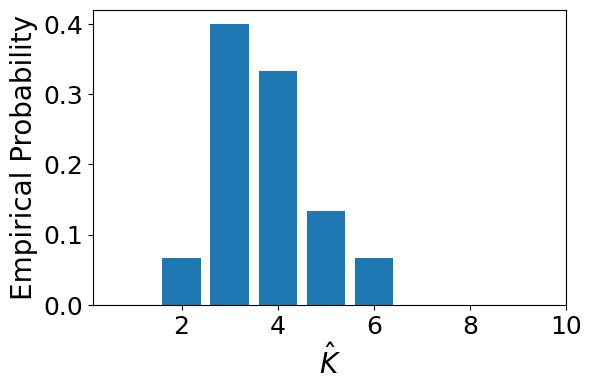}
  \caption{ Simulation results: histogram of $\hat{K}$ under the misspecified case.}
  \label{fig:ms_kselect}
\end{figure}

\begin{figure}[htp]
\setlength{\belowcaptionskip}{-0.cm}
  \centering
  \includegraphics[scale=0.4]{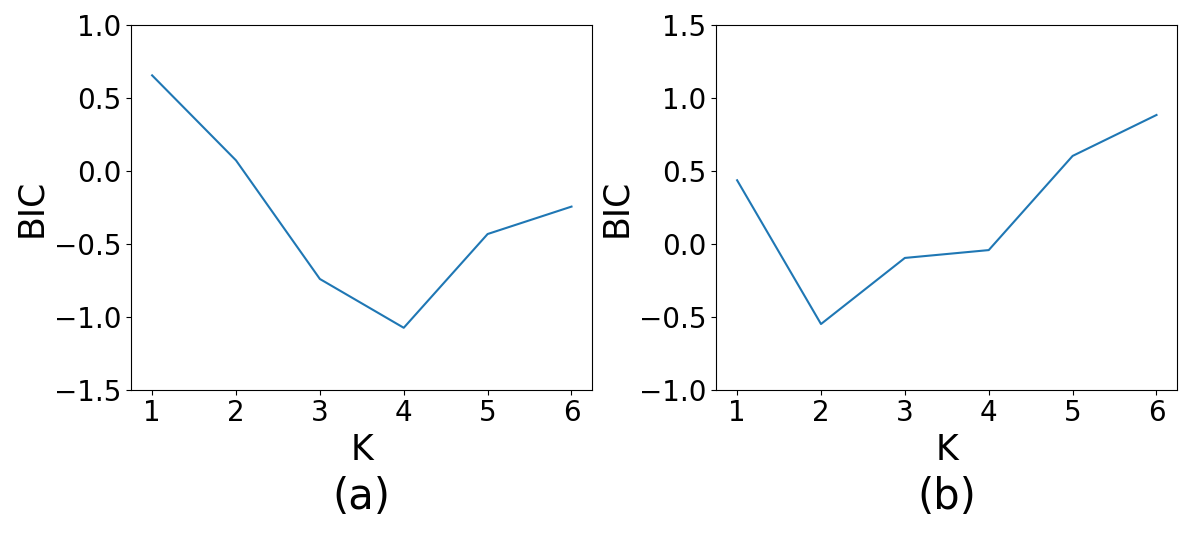}
  \caption{\color{blue}Real data analysis: BIC score under different number of subgroups $K$. (a) CCLE dataset, and (b) TCGA lung cancer dataset.}
  \label{fig:bic_app}
\end{figure}

\begin{figure}[htp]
\setlength{\belowcaptionskip}{-0.cm}
  \centering
  \includegraphics[width=1\textwidth]{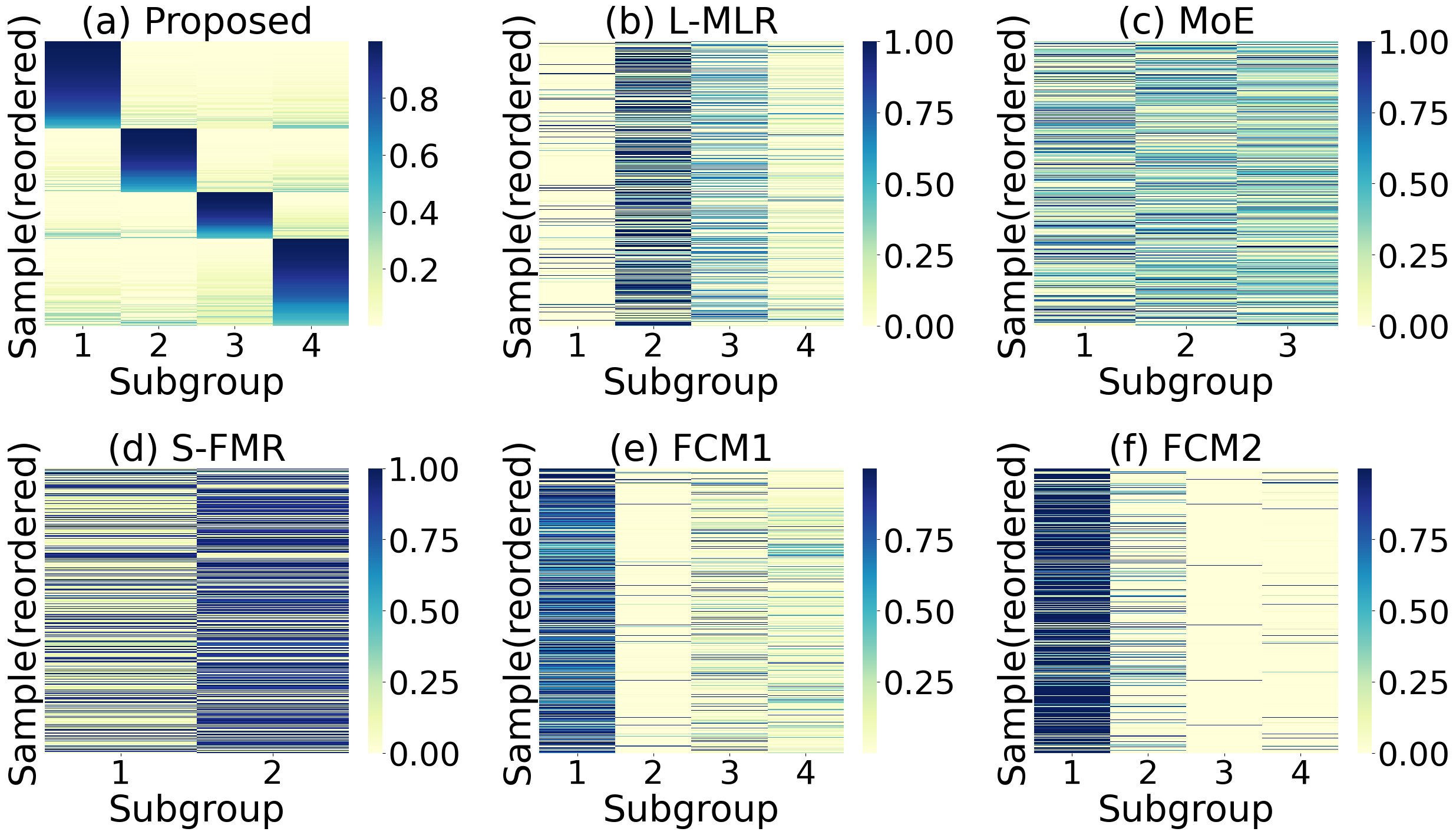}
  \caption{Analysis of CCLE data (response PF2341066): heatmaps of estimated weight matrix. The weights are represented with different colors, as indicated by the colorbar.}
  \label{fig:ccle_group_pf}
\end{figure}

\begin{figure}[htp]
\setlength{\belowcaptionskip}{-0.cm}
  \centering
  \includegraphics[width=0.99\textwidth]{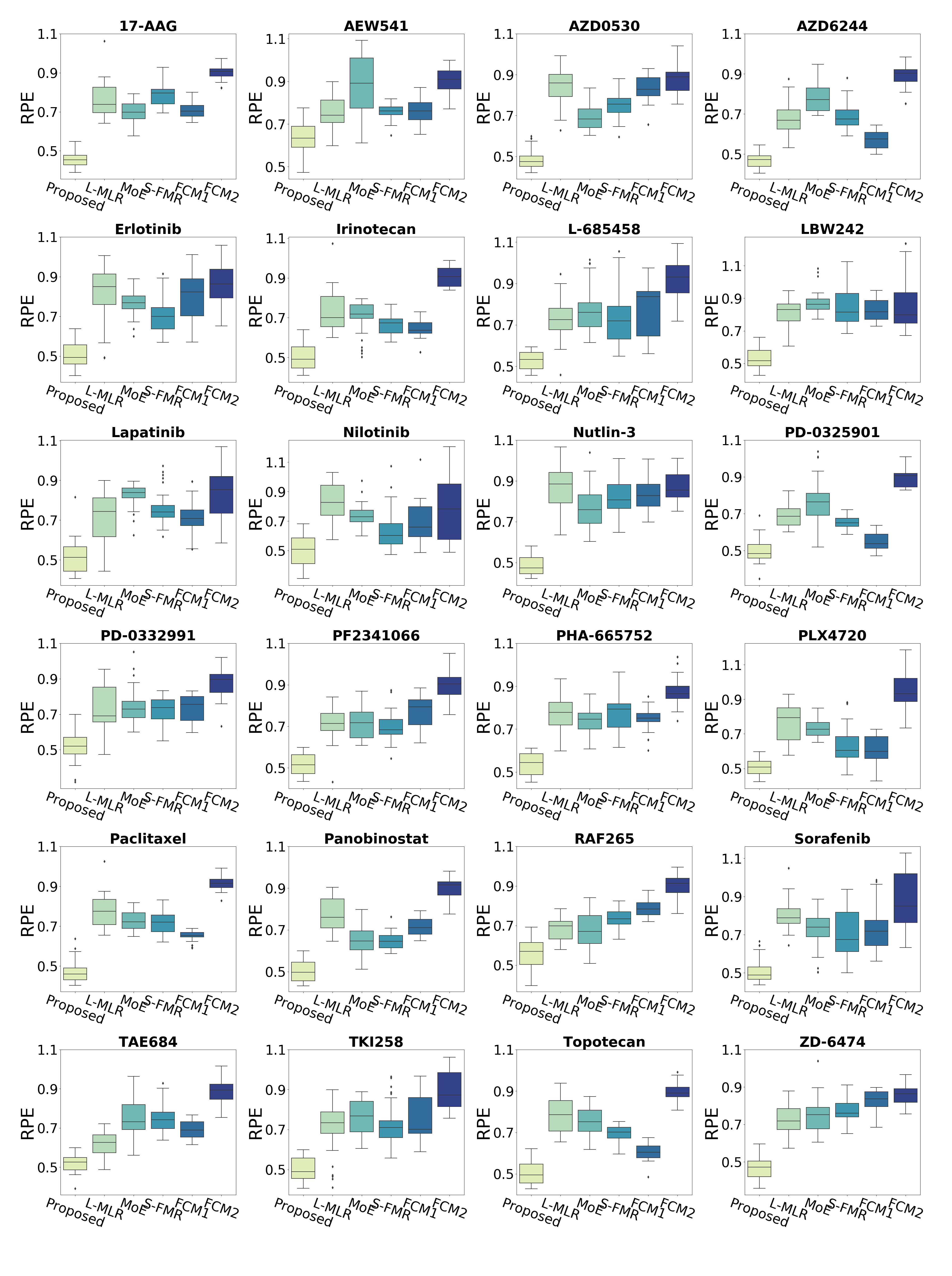}
  \caption{Analysis of CCLE data (response PF2341066): rooted prediction errors for 24 responses (anti-cancer agents).}
  \label{fig:rpe_box_all}
\end{figure}

\begin{figure}[htp]
\setlength{\belowcaptionskip}{-0.cm}
  \centering
  \includegraphics[width=1\textwidth]{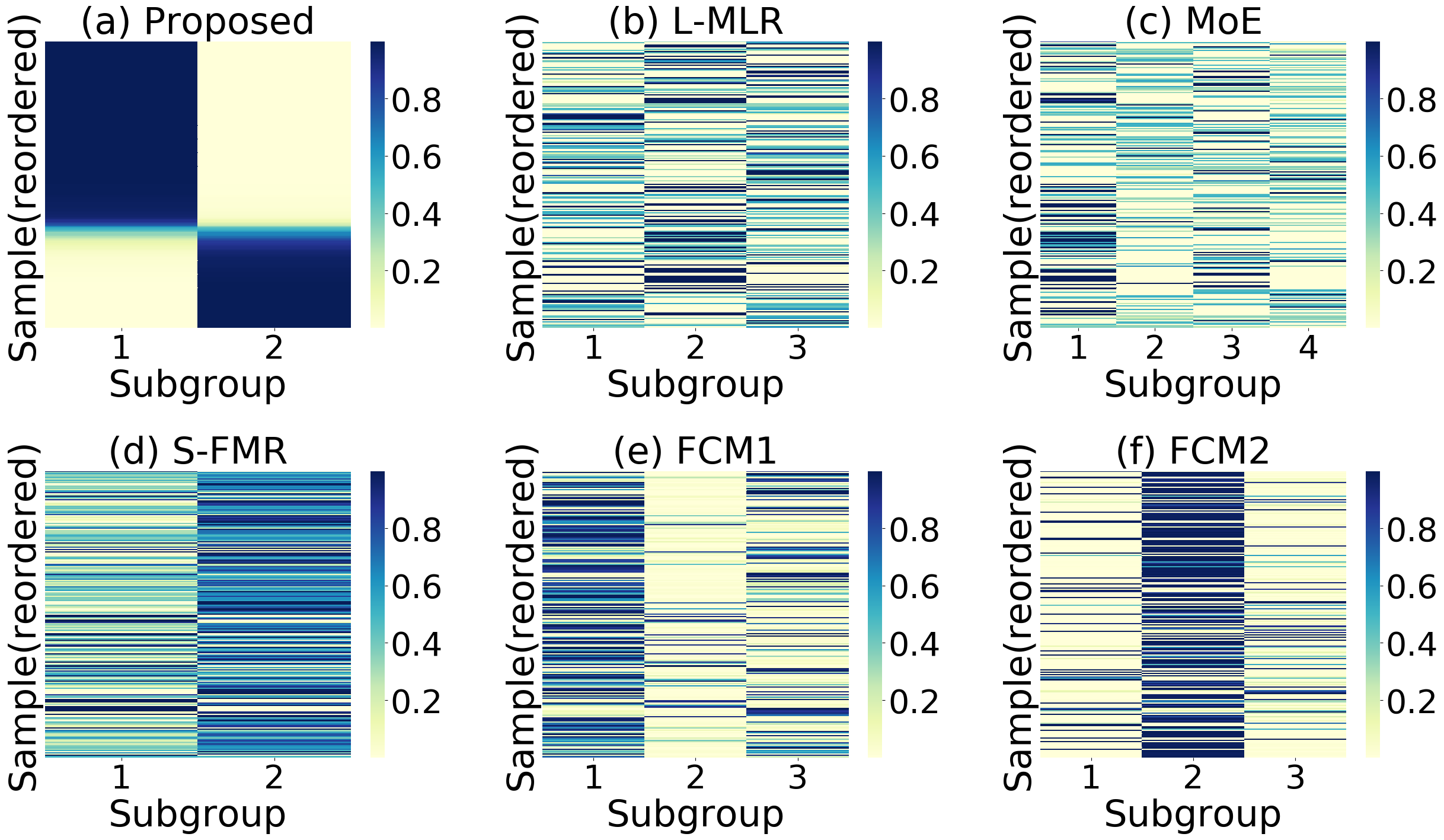}
  \caption{Analysis of TCGA Lung cancer data: heatmaps of estimated weight matrix. The weights are represented with different colors, as indicated by the colorbar.}
  \label{fig:lung_weights_heatmap}
\end{figure}

\clearpage
\section*{D. More details on simulation based on the CCLE data}
The coefficient matrix we consider is 
\[\bm A=\left(
\begin{array}{cccc}
1.5 & 1.5 & 0 & 0\\
-1.5 & 0 & 0 & 0\\
1 & 0 & 0 & 0\\
-1 & 0 & 0 & 0\\
0.5 & 0 & 0 & 0\\
-0.5 & 0 & 0 & 0\\
0.3 & 0 & 0 & 0\\
-0.3 & -1 & 1.5 & 0\\
0 & 0.5 & 0 & 0\\
0 & -0.3 & 0 & 0\\
0 & 0 & -0.5 & 0\\
0 & 0 & 0.3 & 0\\
0 & 0 & 0 & 1\\
0 & 0 & 0 & -1\\
0 & 0 & 0 & 0.5\\
0 & 0 & 0 & -0.5\\
0_{584\times 1} & 0_{584\times 1} &0_{584\times 1} & 0_{584\times 1} \\
\end{array}
\right),\]

and the weight matrix is 

\[\bm U=\left(
\begin{array}{cccc}
1_{102\times 1} & 0_{102\times 1}& 0_{102\times 1}& 0_{102\times 1} \\
0_{102\times 1} & 1_{102\times 1}& 0_{102\times 1}& 0_{102\times 1} \\
0_{102\times 1} & 0_{102\times 1}& 1_{102\times 1}& 0_{102\times 1} \\
0_{102\times 1} & 0_{102\times 1}& 0_{102\times 1}& 1_{102\times 1} \\
a_{1,1} & a_{1,2} &a_{1,3} &a_{1,4}\\
\vdots & \vdots & \vdots & \vdots \\
a_{n-408,1} & a_{n-408,2} & a_{n-408,3} & a_{n-408,4} \\
\end{array}
\right), \]
where $\sum_{j=1}^4 a_{i,j} = 1, \ \ i \in \{1,2,...,n-408\}$ and each $a_{i,j}$ is generated from uniform distribution $\mathcal{U}[0,1]$.

\newpage
\phantom{aaaa}
\end{document}